\documentclass{aa}

\usepackage{graphicx}
\usepackage[varg]{txfonts}

\newcommand{\beq}{\begin{equation}}
\newcommand{\eeq}{\end{equation}}
\newcommand{\bea}{\begin{eqnarray}}
\newcommand{\eea}{\end{eqnarray}}
\newcommand{\req}[1]{Eq.~(\ref{#1})}
\newcommand{\tn}[1]{\mbox{\scriptsize~[#1]}}
\newcommand{\tr}[1]{\tablefootmark{(#1)}}

\newcommand{\dd}{\mathrm{d}}
\newcommand{\ee}{\mathrm{e}}
\newcommand{\gcc}{\mbox{g cm$^{-3}$}}

\begin{document}

\title{Thermal evolution and quiescent emission\\
of transiently accreting neutron stars}

\titlerunning{Thermal evolution of soft X-ray transients}
                                                         
\author{
A. Y. Potekhin\inst{1,2}\thanks{\email{palex@astro.ioffe.ru}}
\and
A. I. Chugunov\inst{2}
\and
G. Chabrier\inst{1,3}
}
\institute{
Ecole Normale
Sup\'erieure de Lyon, CRAL (UMR CNRS 5574), 46 all\'ee d'Italie,
F-69364, Lyon Cedex 07, France
\and
Ioffe Institute,
Politekhnicheskaya 26, 194021, Saint Petersburg, Russia
\and
School of Physics, University of Exeter, Exeter, EX4 4QL, UK
}

\date{Received 03 June 2019 / Accepted 18 July 2019}

\abstract
{}{We study long-term thermal evolution of 
neutron stars in soft X-ray transients
(SXTs), taking the deep crustal heating into account consistently with
the changes of the composition of the crust. We collect 
observational estimates of average accretion rates and thermal
luminosities of such neutron stars and compare the theory with
observations.
}{We perform simulations of thermal evolution of accreting neutron 
stars, considering the gradual replacement of the original
nonaccreted crust by the reprocessed accreted matter, the neutrino and
photon energy losses, and the deep crustal heating due to nuclear
reactions in the accreted crust. We test and compare results for
different modern theoretical models. We  update a compilation of the
observational estimates of the thermal luminosities in quiescence and
average accretion rates in the SXTs and compare the observational
estimates with the theoretical results.
}{Long-term thermal evolution of transiently accreting neutron
stars is nonmonotonic. The quasi-equilibrium temperature in quiescence
reaches a minimum and then increases toward the final steady state.
The quasi-equilibrium thermal luminosity of a neutron star in an SXT
can be substantially lower at the minimum than in the final 
state. This enlarges the range of possibilities for theoretical
interpretation of observations  of such neutron stars. The updates of
the theory and observations leave unchanged the previous conclusions
that the direct Urca process operates in relatively cold neutron stars
and that an accreted heat-blanketing envelope is likely present in
relatively hot neutron stars in the SXTs in quiescence.
The results of the comparison of theory with observations favor 
suppression of the triplet pairing type of nucleon superfluidity in the
neutron-star matter.
}{}

\keywords{stars: neutron -- X-rays: binaries}

\maketitle

\section{Introduction}
\label{sect:intro}

Neutron stars are the most compact stars ever observed: with
typical masses $M\sim 1$\,--\,$2\, M_\odot$, they have radii
$R\approx10-14$ km. The mass density $\rho$ in their core is
$\sim10^{15}$ \gcc, several times normal nuclear density
(the typical density of a heavy atomic nucleus).
Such dense
matter cannot be obtained under laboratory conditions, and its
properties and even composition remain to be clarified.
Since these properties
determine, in particular, the heat loss rate of a neutron star,
a comparison of observed neutron-star surface
luminosities with theoretical predictions is one of the major ways
for studying the extremely dense matter
(see, e.g., \citealt{PPP15} for review and references). 

Many neutron stars reside in binary systems with a lower-mass companion
star (low-mass X-ray binaries, LMXBs) and accrete material onto their
surfaces from the companion. In some cases, the accretion process is
episodic. Such systems, called soft X-ray transients (SXTs), alternate
between phases of accretion (outbursts), lasting usually days to months,
sometimes years, and typically longer periods of quiescence. This
transient activity is regulated most probably by the regime of accretion
from the disks around the neutron stars (e.g., \citealt{Lasota01}). 
During an outburst, the X-ray emission of an LMXB is dominated by the
accretion disk or a boundary layer (e.g., \citealt{InogamovSunyaev10},
and references therein). The released  gravitational energy is so high
that a transient looks like a bright X-ray source with luminosity $\sim
(10^{36}-10^{38})$ erg~s$^{-1}$. During quiescence, the accretion is
switched off or strongly suppressed, and the luminosity decreases by
several orders of magnitude (see, e.g., \citealt{WijnandsDP17} for
review).

In spite of the increase of surface temperature to $\sim10^7$~K during
outburst, inflow of the heat generated by gravitational energy release
is halted due to overheating of deeper layers by nuclear reactions,
associated with compression of the material (e.g.,
\citealt{Fujimoto_ea84,MiraldaEscudeHP90}).

The  neutron star core is predominantly heated by nuclear
reactions occurring in the crust. When the accreted matter falls onto
the neutron star, it pushes the underlying matter down to deeper layers
and thus higher densities, where electron captures, neutron emission,
and pycnonuclear reactions result in the \emph{deep crustal heating},
with the release of $\sim1-2$~MeV per accreted nucleon
\citep{Sato79,HZ90,HZ03,HZ08,Lau_ea18,Fantina_ea18}. Eventually, the
original ground-state ``catalyzed'' crust will be replaced by a crust
composed of accreted matter, while the original crust will have fused
with the core. Once an SXT turns to quiescence, thermal X-ray emission
comes from the surface of the neutron star, so that the thermal
relaxation of the crust can be observed directly (\citealt{Brown_ea98};
see, e.g., \citealt{WijnandsDP17} for a review).

We study the long-term thermal evolution of the SXTs, which determines
the equilibrium level of their quiescent emission. This equilibrium can
be reached after the post-outburst thermal relaxation of the crust, if
the relaxation lasts sufficiently long time. For each neutron star, this
level is a function of the temperature in the stellar core, which is
controlled by the energy losses due to neutrino emission from the core
and the crust and the photon emission from the surface balanced by the
energy income due to the deep crustal heating, which is directly
proportional to the accretion rate. Since the 
time needed for an appreciable heating or cooling of the core
is much longer than the accretion variability
\citep[e.g.,][]{Colpi_ea01,Brown_ea18}, the equilibrium
level is a function of the  \emph{average mass accretion rate}
$\langle\dot{M}\rangle$. Here and hereafter, the angle brackets
$\langle\ldots\rangle$ denote averaging over a timespan covering many
outburst and quiescence cycles. The dependence of the equilibrium
luminosity on $\langle\dot{M}\rangle$ is called \emph{heating curve}
\citep{YakovlevLH03}. Different neutron star models result in different
heating curves, thus giving the means for checking the models by
comparison with observations. 

The heating curves have been previously calculated assuming that the
initial ground-state crust is completely replaced by the reprocessed
accreted matter
\citep{YakovlevLH03,Yakovlev_ea04,BeznogovYakovlev15a,BeznogovYakovlev15b,Fortin_ea18,Matsuo_ea18}.
Meanwhile, it was noted that an amount of the accreted matter might be
insufficient to fill the entire crust
\citep[e.g.,][]{WijnandsDP13,Fantina_ea18}. Here we perform
self-consistent simulations of the long-term thermal evolution, 
considering the gradual replacement of the ground-state crust by an
accreted one and the corresponding evolution of the heat release. We use
a general-relativistic, implicit, adaptive-mesh finite-difference
numerical code, which includes the most recent microphysics input 
(\citealt{PC18}, hereafter Paper~I). We supplement the heating curves
calculated for the fully accreted crust with the analogous curves that
show the position of the minimum of the equilibrium luminosity of the
SXTs on the long-term evolution curves, computed for the constant
average accretion rate. The latter curves, together with the former
ones, enlarge the range of equilibrium luminosities that correspond to a
given average accretion rate and thus increase flexibility of the theory
for explaining the observed equilibrium thermal luminosities of the SXTs
in quiescence. 

For the purpose of comparison of theory to observations, we revisit the
average accretion rates and steady-state thermal luminosities of the
neutron stars in SXTs in quiescence, evaluated from observations.  We
present a list of these properties for 35 SXTs, which includes
relatively new observed SXTs as well as updates of the observational
data traditionally used for comparison with the theoretical heating
curves \citep[e.g.,][]{WijnandsDP17}.

Finally, we explore the effect of suppression of the
nucleon superfluidity by polarization (many-particle correlations),
which is expected to be strong in the case of the triplet pairing gap
(\citealt{Ding_ea16,SedrakianClark}, and references therein). We show
that it brings the theoretical heating curves in full accord with
observations for all transiently accreting neutron stars, including
those whose observed properties were only marginally compatible with the
theory that ignored this effect.

\section{Physics input}
\label{sect:physics}

To model neutron-star cooling processes,  we use the same physics
input as in Paper~I. We briefly summarize it below. 
In this section we also describe the additional ingredients of
the physics input that are brought about by accretion and deep crustal
heating.

\subsection{Equations of thermal evolution}
\label{sect:equations}

In the spherical symmetry, the thermal and mechanical structure of a
star are
governed by six first-order differential equations for radius $r$,
gravitational potential $\Phi$, gravitational mass $M_r$ inside a sphere
of radius $r$, luminosity $L_r$ passing through this sphere, pressure
$P$, and temperature $T$ as functions of the baryon number $a$ interior
to a given shell (\citealt{RichardsonVHS79}; cf.~\citealt{Thorne77}).
Four equations determine the mechanical
structure of the star,
\bea
  \frac{\dd r}{\dd a} &=& \frac{1}{4\pi r^2 \bar{n}}
     \,\left(1-\frac{2GM_r}{r c^2}\right)^{1/2},
\label{drda}
\\
   \frac{\dd M_r}{\dd a} &=& \frac{\rho}{\bar{n}}\,
     \left(1-\frac{2GM_r}{r c^2}\right)^{1/2},
\\
   \frac{\dd\Phi}{\dd a} &=& G\,\frac{M_r+4\pi r^3 P/c^2}{
         4\pi r^4\bar{n}}\,\left(1-\frac{2GM_r}{r c^2}\right)^{-1/2},
\label{Phi}
\\
   \frac{\dd P}{\dd a}&=& -\left(\rho+\frac{P}{c^2}\right)
         \,\frac{\dd\Phi}{\dd a},
\label{dPda}
\eea
where $\bar{n}$ is the mean number density of baryons, $G$ is the
Newtonian constant of gravitation and $c$ is the speed of light in
vacuum. For any known temperature profile $T(a)$ inside the star,
these equations are closed by an equation of state (EoS), which
relates $\rho$ and $P$ to $\bar{n}$ and $T$. In the absence of a strong
magnetic field, we neglect the
dependence of $P$ and $\rho$ on $T$ (use a barotropic EoS) in the inner crust and
the core, but take it into account in the outer crust and envelopes
(Paper~I).

The fifth equation relates the heat flux through a spherical surface 
 to temperature gradient,
\beq
   L_r = -(4\pi r^2)^2\,\bar{n}\,\kappa\,\ee^{-{\Phi/c^2}}\,
        \frac{\dd \ee^{\Phi/c^2}T}{\dd a},
\label{L_T}
\eeq
where $\kappa$ is the thermal conductivity measured in the local
reference frame. 
Finally, time-dependence is introduced by equation
\citep{RichardsonVHS79}
\beq
\frac{\dd (L_r\ee^{2\Phi/c^2})}{\dd a} = 
   \ee^{2\Phi/c^2}\left(\mathcal{E}
   -  T \,\ee^{-\Phi/c^2}
   \frac{\partial s}{\partial t} \right),
\label{Lda}
\eeq
where $\mathcal{E}$ is the net rate of energy generation per baryon and
$\partial s / \partial t$ is the coordinate time derivative of the
entropy per baryon.  The boundary
condition for $\Phi$ is provided by the Schwarzschild metric outside the
star ($r>R$),
\beq
   \ee^{2{\Phi(R)/c^2}} = 1-2GM/c^2R,
\label{PhiR}
\eeq
where $R$ and $M=M_R$ are the stellar radius and mass.

Equations (\ref{drda})\,--\,(\ref{Lda}) assume a spherically symmetric
star in hydrostatic equilibrium. The dynamics of accreted matter is
neglected, because in the case under study  the accretion is extremely
slow on the mechanical relaxation timescales of the concerned layers of
the crust and core of the star.

Equation
(\ref{Lda}) can be combined with \req{L_T} to form
\beq
\frac{c_P}{\bar{n}}\,\ee^{\Phi/c^2}\,
\frac{\partial T}{\partial t} = 
     \frac{\partial}{\partial a}\, \left(4\pi r^2\right)^2\, \bar{n}\, \kappa\,
       \ee^{\Phi/c^2}\,
         \frac{\partial\tilde{T}}{\partial a}
      + \frac{\tilde{Q}}{\bar{n}},
\label{heat_diffusion1}
\eeq
where $c_P$ is the heat capacity per unit volume at constant pressure
and
\beq
  \tilde{Q}=\tilde{Q}_\mathrm{h}-\tilde{Q}_\nu = \bar{n}\, \mathrm{e}^{2\Phi/c^2} \mathcal{E}
\label{tildeQ}
\eeq
is the net heating power density as seen by a distant observer.
Here and hereafter we mark the quantities measured at infinity
(``redshifted'') by tilde
over their symbol. In Eqs.~(\ref{heat_diffusion1}) and (\ref{tildeQ}),
$\tilde{T}=\ee^{\Phi/c^2}T$ and
$\tilde{Q}_{\mathrm{h},\nu}=\ee^{2{\Phi/c^2}}Q_{\mathrm{h},\nu}$, where
$Q_\mathrm{h}$ and $Q_\nu$ are the local heating power and local 
neutrino emission power per unit volume, respectively. 
Assuming that the neutron star is fully
transparent to neutrinos, one can calculate the total heat
release or neutrino emission power in the frame of reference of a
distant observer as (e.g., \citealt{YakovlevLH03})
\beq
   \tilde{L}_{\mathrm{h},\nu} = 
   \int {Q}_{\mathrm{h},\nu}\ee^{2{\Phi/c^2}}\,\dd{V} =
   4\pi\int_0^R 
       \frac{\ee^{2{\Phi/c^2}}Q_{\mathrm{h},\nu}\,r^2\,\dd{r}}{
          \sqrt{1-2GM_r/rc^2}},
\label{tL}
\eeq
where $\dd{V}$ is a proper volume element and the square root in the
denominator is the volume correction factor \citep{Thorne77}.

Equation (\ref{heat_diffusion1}) can be written in the form of  the
usual thermal diffusion equation
\beq
\frac{c_P}{\bar{n}}\,
\frac{\partial\tilde{T}}{\partial t} = 
     \frac{\partial}{\partial a}\, \left(4\pi r^2\right)^2\, \bar{n}\, \kappa\,
       \ee^{\Phi/c^2}\,
         \frac{\partial\tilde{T}}{\partial a}
      + \frac{\tilde{Q}}{\bar{n}}
      + \frac{c_P}{\bar{n}}\,\tilde{T}\,\frac{\partial\Phi}{c^2\partial t}.
\label{heat_diffusion}
\eeq
In practice, the last term on the right-hand side is much smaller than
typical values of the left-hand side.  In Paper~I we treated
it as an external source, with $\partial\Phi/\partial t$ evaluated from
the solution at the preceding time step, but found it insignificant,
therefore we neglect it hereafter. The boundary condition to
\req{heat_diffusion} at the stellar center is
$\partial\tilde{T}/\partial a =0$. The outer boundary condition follows
from \req{L_T} and reads
\beq
\left.
\frac{\partial\tilde{T}}{\partial a}
\right|_{a=a_\mathrm{b}}
 = -\frac{\ee^{{\Phi/c^2}}\,{L}_\mathrm{b}}{(4\pi
r^2)^2\bar{n}\kappa}, 
\label{L_b}
\eeq 
where ${L}_\mathrm{b}$ is the energy flux through the outer boundary
$a=a_\mathrm{b}$, which is provided by the quasi-stationary thermal
structure of a thin envelope outside this boundary. We solve the
nonstationary problem using the temperature-dependent EoS in the outer
crust and choose the mass of the quasi-stationary envelope $\Delta
M_\mathrm{b}$ so as to ensure that plasma is fully ionized at
$\rho>\rho_\mathrm{b}$. In the absence of a strong magnetic field, this
condition is guaranteed for $\Delta M_\mathrm{b}=10^{-12}M_\odot$.

We solve the set of equations (\ref{drda})\,--\,(\ref{L_b}) by a
finite-difference time-implicit scheme with an adaptive mesh and
iterative refinements at each time-step, as described in Paper~I.

\subsection{Equation of state and composition of the core and crust}

There are many theoretical approaches to
construction of the EoS of superdense matter (see, e.g., the review by
\citealt{Oertel_ea17}). For this work we have selected to use just two
most representative models. 

The first model is  BSk24 
\citep{Goriely_CP13,Pearson_ea18}, which provides a unified
treatment of the crust and the core of a neutron star, based on the same
energy-density functional of a modified Skyrme type (so-called Brussels-Montreal functionals). 
It is compatible
with constraints derived from laboratory experiments, and in particular
it ensures the highest accuracy of theoretically computed masses of atomic
nuclei as compared to masses of thousands of different nuclear isotopes
that have been measured in the laboratory. In the stellar core this EoS
is consistent with the EoS of neutron-star core calculated by
\citet{LiSchulze08} within the Brueckner-Hartree-Fock approach, using
the realistic Argonne V18 (Av18) nucleon-nucleon potential \citep{WiringaSC95}
and the phenomenological three-body forces that employ the same
meson-exchange parameters as the Av18 potential.

The second EoS model is A18+$\delta$v+UIX$^*$ \citep{AkmalPR98}, named
APR$^*$ for short.  It is based on variational calculations using
two-body Av18 potential, supplemented by a modified three-body force
UIX$^*$ and so-called relativistic boost interaction (in computations we
use the set of analytical fits to this EoS published in Appendix~A of
Paper~I). The APR$^*$ EoS is applicable only to the
core but not to the crust. In the nonaccreted crust, we supplement it by
the SLy4 EoS \citep{DouchinHaensel01} (for analytic fits, see
\citealt{HP04}).

During accretion, the envelopes, ocean, and crust matter is gradually
replaced by fresh material, whose composition differs from the initial
ground-state matter. In the outer envelopes, up to the density
$\rho\sim10^8-10^9$ \gcc, the initial iron-group element composition is
replaced by the material of the outer layers of the companion star or by
the products of its thermonuclear burning (see \citealt{Meisel_ea18} for
review). For these accreted layers we adopt the layered structure
model of \citet{PCY97} with either H or He on the surface. As soon as
the composition is known, all thermodynamic functions in the outer crust
and the ocean are provided by the analytical model of a fully ionized 
Coulomb plasma \citep{PC13}.

Deeper in the crust, accreted matter is reprocessed by electron
captures, neutron emissions, and pycnonuclear reactions. The
reprocessed matter differs from the exact ground state,
because, for temperature $T\lesssim(4-5)\times10^9$~K, the nuclear
reaction channels relevant for maintaining nuclear statistical
equilibrium (photodissociation, photoabsorption) are closed (see e.g.
\citealt{Bisno01} and references therein). In this case, practical
models determining nuclear composition have been developed by
\citet{HZ90,HZ03,HZ08}. Furthermore, for $T\lesssim3\times10^9$~K,
nuclear shell effects become important and further
contribute to freeze the nuclear composition of the crust. The role of
these effects in the formation of the accreted crust has been recently
studied by \citet{Fantina_ea18}. In the present study, we use and
compare two most recent models \citep{HZ08,Fantina_ea18}, described
below.

\subsection{Heat loss and production} 

Cooling of an isolated neutron star goes through two major stages. The
first, \emph{neutrino cooling} stage lasts $\sim10^5$ years. During this
period, the core cools mostly via neutrino emission. The second,
\emph{photon cooling} stage begins when, with temperature decrease, the
neutrino energy losses become smaller than the losses due to
electromagnetic radiation from the surface.
 However, an accreting neutron star may become
sufficiently hot again and return to the neutrino
cooling regime.

\citet{YKGH} presented a comprehensive review of neutrino emission
mechanisms in compact stars and supplied convenient fitting formulae for
astrophysical applications. The most important reactions in the
neutron-star crust and core with references to the appropriate fitting
formulae are collected in Table~1 of \citet{PPP15}, which also includes
references to several important updates in the relevant neutrino
reaction rates, which improve the results of \citet{YKGH}.

Most powerful neutrino emission occurs in the direct Urca process, but
it operates only if the proton fraction $Y_\mathrm{p}$ exceeds some
threshold value $Y_\mathrm{pDU}$, which occurs above a certain threshold
baryon number density $\bar{n}_\mathrm{DU}$. In the
neutron-proton-electron ($npe$) matter $Y_\mathrm{pDU}=1/9$, but in the
$npe\mu$ matter (with allowance for $\mu^-$-mesons) it is generally
larger \citep[e.g.,][]{Haensel95}. The proton fraction as a function of
the mean baryon number density $\bar{n}$ is uncertain, since it  depends
on the microscopic interaction model. A self-consistent modeling should
employ the same interaction model for the EoS and for the proton
fraction calculations. In this case, the threshold density depends on
the EoS. Specifically, for the BSk24 and APR$^*$ models we have
$\bar{n}_\mathrm{DU}=0.0453$ fm$^{-3}$ ($Y_\mathrm{pDU}=0.136$) and
$\bar{n}_\mathrm{DU}=0.0783$ fm$^{-3}$ ($Y_\mathrm{pDU}=0.141$),
respectively. These densities are reached only in central parts of
sufficiently massive neutron stars. The minimal mass of the star that
allows the direct Urca processes to operate is
$M_\mathrm{DU}=1.595\,M_\odot$ for BSk24 and
$M_\mathrm{DU}=2.01\,M_\odot$ for APR$^*$. In the absence of the direct
Urca processes, the most important neutrino emission mechanisms in the
core are the modified Urca (Murca) processes, baryon bremsstrahlung, and
Cooper pair breaking and formation whenever the baryons are superfluid.

\begin{figure}
\centering
\includegraphics[width=\columnwidth]{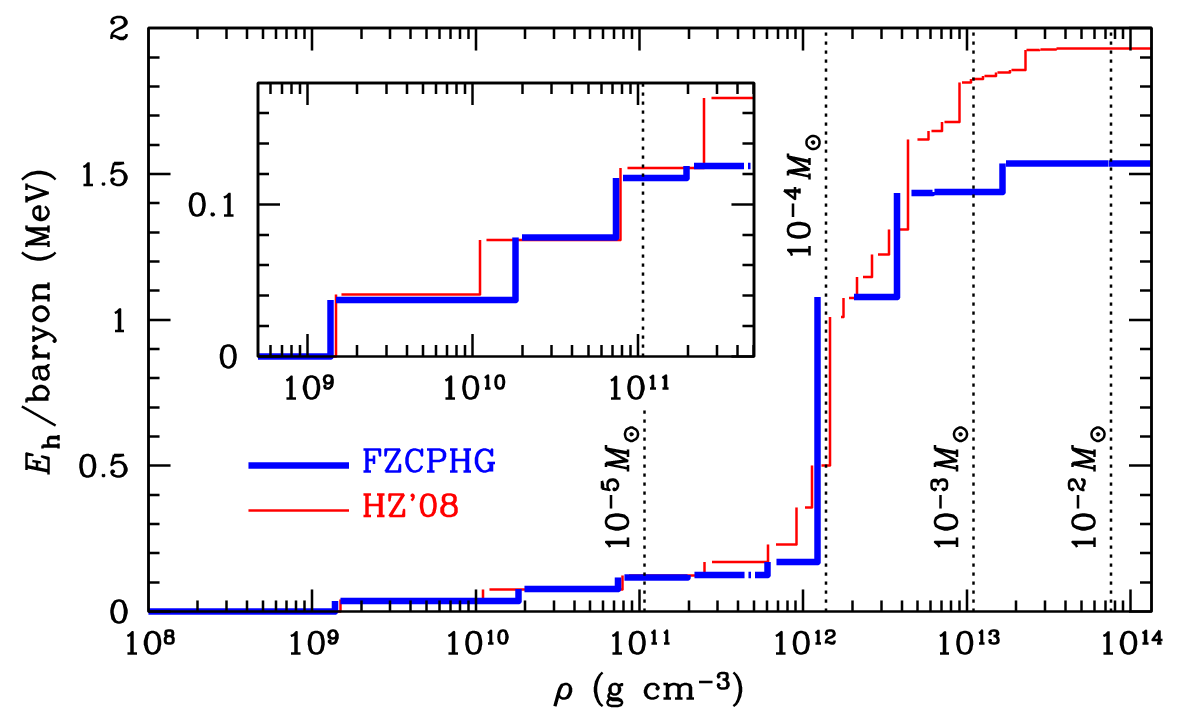}
\caption{Total heat $E_\mathrm{h}$ generated per an accreted baryon, as
function of mass density $\rho$, according to the models of
\citet{Fantina_ea18} (FZCPHG, thick blue line) and \citet{HZ08} (HZ'08,
thinner red line). The gaps in the lines correspond to the density
discontinuities at the phase boundaries.
The vertical dotted lines mark the $\rho$ values
corresponding to four masses of accreted material, from
$10^{-5}\,M_\odot$ to $10^{-2}\,M_\odot$, labeled near these lines,
for a neutron star with
gravitational mass $M=1.4\,M_\odot$ and radius $R=12.6$~km. The inset
shows a zoom to the low-density region.
}
\label{fig:heatsrc}
\end{figure}

The nuclear transformations in the crust during accretion are
accompanied with energy release. Part of this energy is lost to neutrino
emission, but another part is transformed in heat which warms up the
stellar crust.  Here we consider two most recent models of the deep
crustal heating, developed without and with allowance for the nuclear
shell effects, respectively, by \citet{HZ08} (HZ'08) and by
\citet{Fantina_ea18} (FZCPHG). Each of these two models has several
versions. For the first model (without nuclear shell effects), we choose
the version of the accreted-crust composition and respective energy
releases at the boundaries of different layers that is given in
Table~A.3\footnote{We have fixed a typo (``0.8'' should read ``0.08'')
for the density discontinuity at $\rho=1.766\times10^{12}$ \gcc{}.} of
\citet{HZ08}. For the second one (with the shell effects) we adopt the
results reported in Table~A.1 of \citet{Fantina_ea18}. Both tables
correspond to the initial iron composition. The first table (HZ'08) is
based on compressible liquid drop model by \citet{MB77}, while the
second table (FZCPHG) corresponds to the BSk21  energy-density
functional model (\citealt{Goriely_CP10}), which is similar to the BSk24
model that underlies the basic EoS used in the present work for the
ground-state matter. The HZ'08 model predicts a total release of 1.93
MeV of heat per accreted baryon, and the FZCPHG model predicts 1.54 MeV
per baryon. Figure~\ref{fig:heatsrc} displays the total heat generated
per accreted baryon, from the surface to a given density in the crust,
as function of mass density. The vertical dotted lines correspond to
several masses of accreted material, from the surface to the given
density, for a neutron star with gravitational mass $M=1.4\,M_\odot$ and
radius $R=12.6$~km, consistent with the BSk24 EoS.

\subsection{In-medium effects} 

Neutrino emissivity of neutron stars can be strongly modified by
in-medium (collective) effects, which affect the
reaction rates in several ways \citep[see][for a
review]{Voskresensky01}. Their simplest manifestation is the
modifications of the effective nucleon masses $m_\mathrm{p}^*$ and
$m_\mathrm{n}^*$ owing to distortion of the dispersion relation. The
values of these effective masses should be taken from microscopic
theories. The ratios $m^*/m$ affect not only the neutrino emission
rates, but also baryon thermal conductivities, and thus they have a
complex effect on thermal evolution of the star.  The ratios
$m_\mathrm{n}^*/m_\mathrm{n}$ and  $m_\mathrm{p}^*/m_\mathrm{p}$
(respectively, for neutrons and protons), used in the present work, are
plotted in the bottom panel of Fig.~\ref{fig:meff_Tc_Y} as functions of
the mean baryon density $\bar{n}$, according to the microscopic theories
consistent with the two equations of state that we employ in this work.
For BSk24, they are given by Eq.~(A10) of \citet{Chamel_GP09} with the
parameters listed in \citet{Goriely_CP13}. For APR$^*$, we use Eq.~(6)
of \citet{Baldo_ea14} with the parameters  for the effective two- and
three-body potentials Av18+UIX that underlie the EoS APR$^*$. The number
densities of the free nucleons of each type, which are needed in these
equations, are calculated according to the fitting formulas given in
\citet{Pearson_ea18} for BSk24 and in Paper~I for APR$^*$.

However, the effective mass approximation may be insufficient, being
unable to describe some qualitative in-medium effects that are absent in
the free space. The nucleon correlations also affect reaction matrix
elements and propagators and modify the density of intermediate states.
In particular, according to \citet{Schaab_ea97}, the
in-medium effects enhance emissivity in the Murca process and result in
a strong density dependence, which gives a smooth crossover from the
standard  to the enhanced cooling scenario for increasing star masses. A
qualitatively similar effect was found by \citet{ShterninBH18}, who
described the in-medium nucleon scattering in the
Brueckner-Hartree-Fock approximation taking the effective
two- and three-body forces and the Pauli blocking of intermediate
states into account. These authors suggested a simple expression for the
medium-enhanced emissivity of the neutron branch of the Murca process,
which has been incorporated in our code. The importance of this
enhancement of the Murca process for neutron star cooling was
demonstrated by \citet{ShterninBH18} and in Paper~I.

\subsection{Baryon superfluidity} 
\label{sect:SF}

Baryon superfluidity is known to affect neutron star thermal evolution,
first, due to its influence on the  heat capacity, neutrino emissivity,
and heat transport, and second, due to the emergence of a
specific neutrino emission mechanism by Cooper pair breaking and
formation (PBF) (see, e.g., the reviews by \citealt{Page_ea13} and by
\citealt{SchmittShternin18}). The PBF processes are most powerful at
$T\sim T_{\mathrm{crit}}$, where $T_\mathrm{crit}$ is a critical
temperature, specific for each type of superfluidity \citep[e.g.][and
references therein]{Leinson10}. Microscopic theories and methods that
are being employed to understand the basic properties of superfluid
nuclear systems, with emphasis on the neutron-star matter, have been
recently reviewed by \citet{SedrakianClark}. 

To incorporate these effects in astrophysical modeling, we use the
convenient fitting formulas collected by \citet{YKGH} with updates and
corrections listed in \citet{PPP15}. As a rule, these fitting
formulas describe the effects of superfluidity as functions of
$T/T_\mathrm{crit}$, where  the critical temperature $T_\mathrm{crit}$
depends on the nucleon type (neutrons or protons) and their Cooper
pairing type. For each type of superfluidity, $T_\mathrm{crit}$ also
depends on the number density of free nucleons. Different theoretical
results for these dependencies have been parametrized by
\citet{Ho_ea15}. In the present work, we consider the parametrizations
that describe theoretical results of \citet{BaldoSchulze07} (BS) for
proton singlet ($^1S_0$) pairing type, \citet{Baldo_ea98} (BEEHS) for
neutron triplet ($^3P_2-^3F_2$) pairing type, and either
\citet{MargueronSH08} (MSH) or \citet{Gandolfi_ea09} (GIPSF) for neutron
singlet ($^1S_0$) pairing type. The first two types of superfluidity are
most relevant in the core of a neutron star, and the last one in the
crust.

\begin{figure}
\centering
\includegraphics[width=\columnwidth]{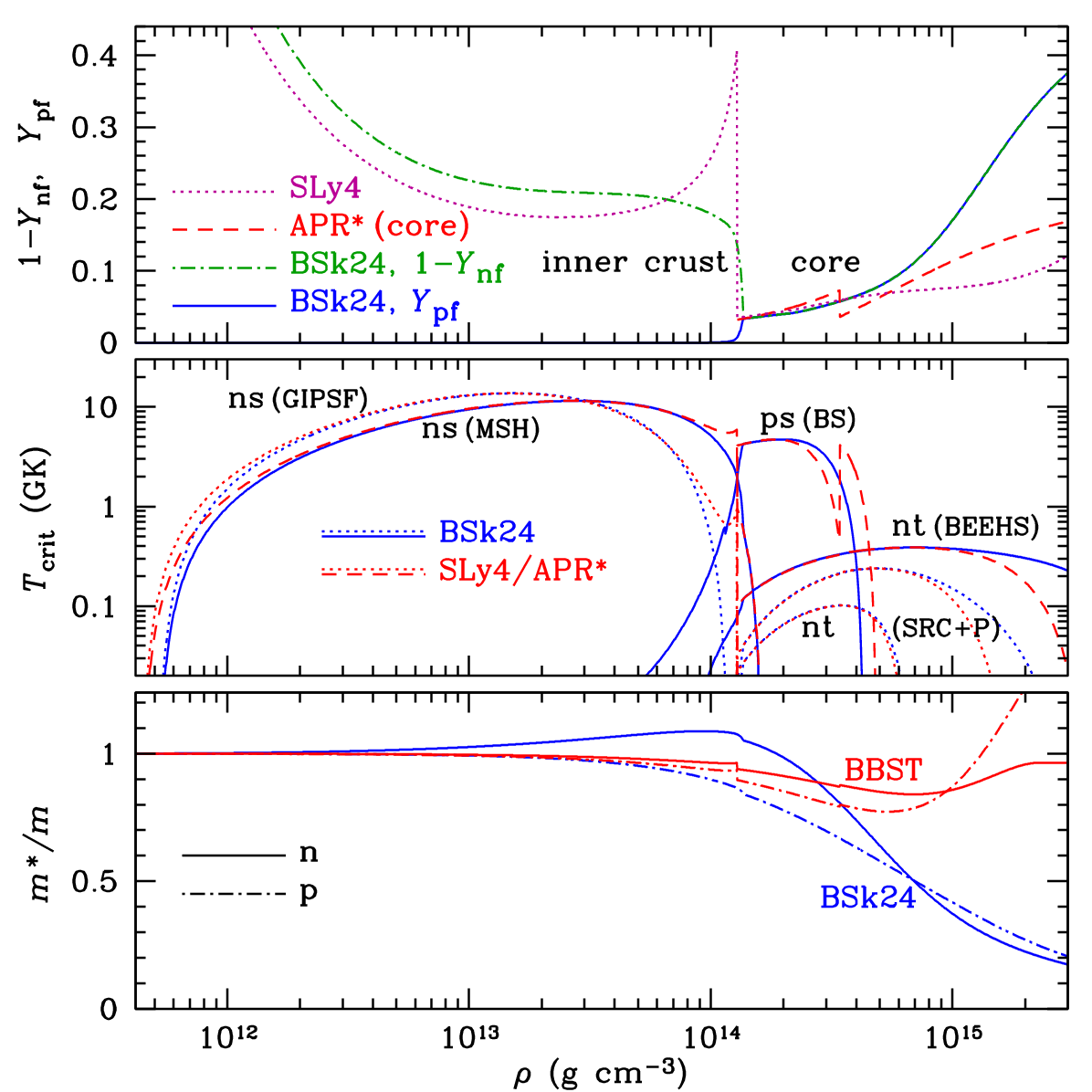}
\caption{Free nucleon fractions (top panel), critical temperatures
(middle panel), and relative effective masses (bottom panel) for the
basic theoretical models used in this paper for the nonaccreted crust
and the core of a neutron star. The unified generalized Skyrme model
BSk24 \citep{Pearson_ea18} is compared with the variational APR$^*$
model for the core \citep{AkmalPR98}, and with the SLy4 model
\citep{DouchinHaensel01} for the crust EoS and composition, as well as with the
results of \citet{Baldo_ea14} (BBST) for the effective masses. For the
critical temperatures $T_\mathrm{crit}$  of proton singlet (ps),
neutron singlet (ns), and neutron triplet (nt) pairing
types of superfluidity we employ the results of
\citet{BaldoSchulze07} (BS), \citet{MargueronSH08} (MSH),
and \citet{Baldo_ea98} (BEEHS), respectively, as parametrized by
\citet{Ho_ea15}. For comparison, by dotted lines in the middle panel we
show the results  of \citet{Gandolfi_ea09} (GIPSF) for the
ns-superfluidity in the neutron star crust, also parametrized by
\citet{Ho_ea15}, and the results of \citet{Ding_ea16} for the
nt-superfluidity in the neutron star core, calculated  with allowance
for short-range correlations and polarization (SRC+P) for the effective
potential models N3LO (upper dotted curves) and Av18 (lower dotted
curves).
}
\label{fig:meff_Tc_Y}
\end{figure}

The top panel of Fig.~\ref{fig:meff_Tc_Y} shows number fractions of free
neutrons ($Y_\mathrm{nf}$) and protons ($Y_\mathrm{pf}$) as functions of
mean baryon number density $\bar{n}$ in the inner crust and the core of
a neutron star for the EoS models described above. Corresponding values
of $T_\mathrm{crit}$ as functions of $\bar{n}$ are shown in the middle
panel for the above-mentioned theoretical models of baryon pairing
gaps. 

Recent studies have demonstrated that the effects of many-body
correlations on baryon superfluidity can suppress the superfluid gap,
and consequently $T_\mathrm{crit}$, by an order of magnitude or even
stronger for the triplet type of pairing $^3P_2-^3F_2$ (e.g.,
\citealp{Ding_ea16}; see \citealt{SedrakianClark} for a discussion). The
suppressed critical temperatures are also shown in the middle panel of
Fig.~\ref{fig:meff_Tc_Y} (below the BEEHS curves). The influence of this
effect on the cooling of isolated neutron stars has been recently
studied by \citet{WeiBS19}. In Sect.~\ref{sect:nt_suppr} we test
its influence on the quiescent thermal states of neutron stars in the
SXTs.

\section{Equilibrium thermal luminosities in quiescence} 

\subsection{Observations} 

During a long period covering many outbursts, the interior of a neutron
star in an SXT becomes appreciably heated by the part of the deep
crustal heat that flows into the core during accretion. The temperature
of the core $T_\mathrm{core}$ thus increases, until this heating is
balanced by the neutrino energy loss. The higher is the average mass
accretion rate $\langle \dot{M}\rangle$, the higher is the equilibrium
$T_\mathrm{core}$ value at the crust-core boundary. This value
determines a thermal equilibrium state that the crust tends to acquire
in quiescence. Thus thermal photon luminosity of the SXT in quiescence,
$L_\mathrm{q}$, is correlated with $\langle \dot{M}\rangle$. A concrete
value of $L_\mathrm{q}$ at a given $\langle \dot{M}\rangle$ depends on
the neutron star parameters and on the properties of the dense matter in
the interior of the star.
It also depends on the properties of the heat blanketing
envelopes: if the accreted matter has been burnt to heavy chemical
elements (one usually takes iron for a fiducial model), the thermal
luminosity is lower than in the cases where the heat transport is
controlled by layers composed of relatively light chemical elements,
for instance if the mass of residual helium is sufficiently large to
fill the heat blanket (see, e.g., \citealt{Yakovlev_ea04}, and
references therein). Thus the simultaneous consideration of $\langle
\dot{M}\rangle$ and $L_\mathrm{q}$ can help to determine neutron-star
parameters and probe the properties of neutron star interior.

\citet{YakovlevLH03} were the first to undertake such study. They
considered five SXTs whose average accretion rate and quiescent
luminosity had been estimated by that time. More comprehensive
compilations of the properties of the SXTs in quiescence were published
by \citet{Heinke_ea07,Heinke_ea09,Heinke_ea10}. Thereafter, these data,
for 24 SXTs in total, have been traditionally  quoted  in different
reviews and research papers (e.g.,
\citealt{WijnandsDP13,WijnandsDP17,BeznogovYakovlev15a,BeznogovYakovlev15b})
for analysis of the $L_\mathrm{q}(\langle\dot{M}\rangle)$
correlations. Some sources beyond this sample were discussed in the
context of such analysis from time to time (for example, 1RXS
J180408.9$-$342058 has been added to the sample by
\citealt{Parikh_ea18}, SAX J1750.8--2900 was discussed by
\citealt{Lowell_ea12} and by \citealt{ParikhWijnands17} and plotted in a
figure by \citealt{Fortin_ea18}), and the average mass transfer rates
for some of the SXTs  have been revisited
\citep{CoriatFD12,Heinke_ea13,VanIH19}, but a systematic revision of the
cumulative dataset has not been undertaken.

Here we present a renewed and more comprehensive  compilation of the
observational data pertinent to the analysis of the quasi-equilibrium
thermal states of neutron stars in the SXPs in quiescence.
Tables~\ref{tab:objlist} and~\ref{tab:obj} present 35 objects, suitable for the analysis of
the $L_\mathrm{q}$($\langle\dot{M}\rangle$) dependence. We have not only
expanded the list, but also updated the average accretion rates and/or
quiescent luminosities for most of the previously tabulated 24 SXTs. In
order to preserve continuity with the previous works, the SXTs numbered
1--15 in our tables are the same objects as in
\citet{Heinke_ea10} and \citet{WijnandsDP17}, while our SXT numbers 16
through 24 had been previously labeled by letters A through I,
respectively.

\begin{table}
\centering
\caption[]{List of SXTs 
with estimated average accretion rates
and quiescent thermal luminosities.
The first column gives the sequential number for a quick reference; 
the second column lists the most common source identifiers in the literature;
and the last column indicates a particular source type
or association.
\label{tab:objlist}}
\begin{tabular}{rlc}
\hline\hline
\noalign{\smallskip}
 no.&  Source                                  &  Remark\tr{a}       \\
\hline                                                                                                                                                                            
\noalign{\smallskip}                                                                                                                                                              
  1 &  4U 2129+47 (V$^\ast$ V1727 Cyg)         &  In NGC 7078        \\
  2 &  KS 1731$-$260                           & CC, QP              \\
  3 &  4U 1608$-$522                           &                     \\
  4 &  EXO 1745$-$248 (Ter 5 X-1)              &        In Terzan 5  \\
  5 &  1M 1716$-$315 (1H 1715-321)             &                     \\
  6 &  RX J1709.5$-$2639                                             \\
    &~~~~(XTE J1709$-$267)                     &     In NGC 6293     \\
  7 &  MXB 1659$-$29 (XB~1658$-$298)           & CC, QP, E           \\
  8 &  1RXS J173546.9$-$302859                                       \\
    & ~~~~(XB 1732$-$304)                      &        In Terzan 1  \\
  9 &  4U 1456$-$32 (Cen X-4)                  &                     \\
 10 &  1H 1905+00 (4U 1857+01)                 &                     \\
 11 &  SAX J1806.8$-$2435 (2S 1803$-$245)      &                     \\
 12 &  4U 1730$-$22                            &                     \\
 13 &  EXO 1747$-$214                          &                     \\
 14 &  XTE 2123$-$058                          &                     \\
 15 &  SAX J1810.8$-$2609                      &                     \\
 16 &  4U 1908+005 (Aql X-1)                   &  CC                 \\
 17 &  SAX J1748.9$-$2021 (NGC 6440 X-1)       &  In NGC 6440        \\
 18 &  CXOGlb J174852.7$-$202124               &   UC                \\
    &~~~~(NGC 6440 X-2)                        &  in NGC 6440        \\
 19 &  XTE J0929$-$314   (V$^\ast$ BW Ant)     &   UC                \\
 20 &  SAX J1808.4$-$3658 (V$^\ast$ V4580 Sgr) &                     \\
 21 &  XTE J1807$-$294                         &   UC                \\
 22 &  XTE J1751$-$305                         &   UC                \\
 23 &  XTE J1814$-$338 (V$^\ast$ V5511 Sgr)    &                     \\
 24 &  IGR J00291+5934 (V$^\ast$ V1037 Cas)    &                     \\
 25 &  HETE J1900.1$-$2455                     & CC, QP              \\
 26 &  XTE J1701$-$462                         & CC, QP              \\
 27 &  IGR J17480$-$2446                       & CC                  \\
    &~~~~(Ter 5 X-2)                           & in Terzan 5         \\
 28 &  EXO 0748$-$676  (V$^\ast$ UY Vol)       & CC, QP, E           \\
 29 &  1RXS J180408.9$-$342058                 & CC                  \\
 30 &  Swift J174805.3$-$244637                & CC                  \\
    &~~~~(Ter 5 X-3)                           &   in Terzan 5       \\
 31 &  SAX J1750.8$-$2900                      &                     \\
 32 &  Swift J1756.9$-$2508                    & UC                  \\
 33 &  Swift J1750.7-3117                      &        E            \\
    &~~~~(GRS 1747$-$312)                      &   in Terzan 6       \\
 34 &  IGR J18245$-$2452                       & In Messier 28  \\
 35 &  MAXI J0556$-$332                        &       CC, QP        \\
\noalign{\smallskip}                                                                                                                                     
\hline\hline
\end{tabular}
\tablefoot{
\tablefoottext{a}{UC -- ultra-compact source \citep{VanIH19},
 E -- eclipser, QP -- quasi-persistent source,
 CC -- crust cooling source \citep{WijnandsDP17}.}
}
\end{table}

\begin{table*}
\centering
\caption[]{Key properties of SXTs 
with estimated average accretion rates
and quiescent thermal luminosities.
Each row gives the sequential number; 
source name (may be truncated; see Table~\ref{tab:objlist} for the full identifiers);
estimate of the long-term averaged mass accretion rate; 
observed thermal luminosity in quiescence;
orbital period;
estimate of the companion (donor star) mass (some mass
ratio estimates are given in the footnotes);
distance estimate.
\label{tab:obj}}
\begin{tabular}{rlccllcc}
\hline\hline
\noalign{\smallskip}
 no.&  Short name       &$\langle\dot{M}_\mathrm{obs}\rangle$& $\tilde{L}_\mathrm{q}$            & $P_\mathrm{orb}$& spin          &         $M_\mathrm{d}$      & Distance                   \\
    &                   & ($M_\odot$ yr$^{-1}$)              & ($10^{33}$ erg s$^{-1}$)          &  (h)            &  (Hz)         &         ($M_\odot$)         &   (kpc)                    \\
\hline                                                                                                                                                              
\noalign{\smallskip}                                                                                                                                                
  1 &  4U 2129+47       & $3.9\times10^{-9 }  $\tn{1}      &  $1.5^{+3.1}_{-1.2}$\tn{2,3}\tr{a}  & 5.96\tn{1}      & --            &   --                        & $ 10.3\pm0.4       $\tn{4}  \\
  2 &  KS 1731$-$260    & $<9\times10^{-10}$\tn{5}\tr{b}&  $0.39\pm0.03$\tn{6}                   &  $>2$\tn{7}     & 524\tn{8}     &   --                        & $ 7.2\pm1.0        $\tn{9}  \\
  3 &  4U 1608$-$522    & $9.6\times10^{-10}  $\tn{1}      &  $5.3^{+4.7}_{-2.9}$\tn{10,11}\tr{a}&  10\,--125\tn{7}& 620\tn{8}     &   --                        & $ 4.1\pm0.4        $\tn{9}  \\
  4 &  Ter 5 X-1        & $3\times10^{-11}    $\tn{12}     &  $<0.1 $\tn{12,13}                  &    --           & 333\tn{14}    &   --                        & $ 5.5\pm0.9        $\tn{15} \\
  5 &  1M 1716$-$315    & $<2.5\times10^{-10} $\tn{2}      &  $1.3^{+1.2}_{-0.7}$\tn{2,16}\tr{a} &    --           & --            &   --                        & $ 5.1-6.9          $\tn{17} \\
  6 &  XTE J1709        & $1.8\times10^{-10}  $\tn{10}      & $1.4^{+0.6}_{-0.5}$\tn{18}\tr{b}    &   --            & --            &   --                        & $ 8.5\!-\!8.8      $\tn{18,19} \\
  7 &  MXB 1659$-$29    & $1.7\times10^{-10}  $\tn{10}      &$0.20^{+0.05}_{-0.11}$\tn{10,20}\tr{a}&   7.11\tn{9}    & 567\tn{8}     &  0.3--0.8\tn{21}            & $ 12\pm3           $\tn{9}  \\
  8 &  XB 1732$-$304    & $<1.5\times10^{-10} $\tn{10}      &  $<1.1$\tn{10}                       &    --           &  --           &   --                        & $ 5.2\pm0.5        $\tn{22} \\
  9 &  Cen X-4          & $3.8\times10^{-11}  $\tn{1}      &  $0.12\pm0.01$\tn{23}\tr{b}         &   15.1\tn{24}   &  --           & $0.31\pm0.27$\tn{25}\tr{c}  & $ 1.2\pm0.2        $\tn{1}  \\
 10 &  1H 1905+00       & $<1.1\times10^{-10} $\tn{2}      &  $<0.01$\tn{2}                      &    $<1.5$?\tn{2}&  --           &   --                        & $ 10        $\tn{26}\tr{d} \\
 11 &  2S 1803$-$245    & $<7\times10^{-11}   $\tn{2}      &  $<0.52$\tn{2}                      &   $\sim9$?\tn{2}&  --           &   --                        & $ 7.3    $\tn{2,27}\tr{d}  \\
 12 &  4U 1730$-$22     & $<4.8\times10^{-11}$\tn{2}       &  $2.2^{+2.0}_{-1.1}$\tn{2,28}\tr{a} &    --           &  --           &   --                        & $ 10^{+12}_{-4}    $\tn{28} \\
 13 &  EXO 1747         & $<3\times10^{-11}   $\tn{10}      &  $<0.07$\tn{10}                      &    --           &  --           &   --                        & $ 11     $\tn{9,29}\tr{d}  \\
 14 &  XTE 2123         & $<7\times10^{-12}   $\tn{1}      &  $<0.14$\tn{10}                      &    5.956\tn{7}  &  --           & $0.76\pm0.22$\tn{30}\tr{e}  & $ 9.6\pm1.3        $\tn{30} \\
 15 &  SAX J1810.8      & $5\times10^{-12}    $\tn{31}     &  $<0.2$\tn{10}                       &    --           &   532\tn{32}  &  --                         & $ 4.9\pm0.3        $\tn{33} \\
 16 &  Aql X-1          & $3.2\times10^{-10}$\tn{34}\tr{b} &  $2.1\pm0.5$\tn{34}\tr{b}           &   18.9\tn{7}    &   550\tn{7}   &  --                         & $3.0-6.1           $\tn{7}  \\
 17 &  NGC 6440 X-1     & $6\times10^{-11}$\tn{35}         &  $1.3\pm0.4$\tn{36}\tr{b}           &   8.765\tn{37}  &   442\tn{37}  &  0.12--1\tn{37}             & $ 8.5\pm0.6        $\tn{37} \\
 18 &  NGC 6440 X-2     & $8.4\times10^{-13}$\tn{35}       &  $<0.023$\tn{38}\tr{b}              &   0.960\tn{39}  &   206\tn{39}  & $\sim0.0076$\tn{39}         & $ 8.5\pm0.4        $\tn{40} \\
 19 &  XTE J0929        & $\lesssim2\times10^{-11}$\tn{2}  &  $<0.1$\tn{2}\tr{f}                 &   0.726\tn{41}  &   185\tn{41}  & $\sim0.01$\tn{41}           & $ 8^{+7}_{-3}      $\tn{40} \\
 20 &  SAX J1808.4      & $1.7\times10^{-11}$\tn{1}        &  $<0.02$\tn{2}\tr{f}                &     2.014\tn{42}& 401\tn{42}    &$0.04^{+0.02}_{-0.01}$\tn{43}& $   3.4-3.6        $\tn{1}  \\
 21 &  XTE J1807        &$\lesssim3\times10^{-11}$\tn{35}\tr{g}&     $<0.13$\tn{2}               & 0.667\tn{44}    &   190.6\tn{44}&   --                        & $ 4.4\pm0.6        $\tn{44} \\
 22 &  XTE J1751        & $4.3\times10^{-12}  $\tn{35}     &  $<0.4$\tn{2}                       &   0.707\tn{45}  &    245\tn{8}  & 0.014--0.035\tn{45}         & $ 8^{+0.5}_{-1.3}  $\tn{40} \\
 23 &  XTE J1814        & $6\times10^{-12}    $\tn{1}      &  $<0.17$\tn{2}                      &     4.275\tn{7} & 314\tn{7}     & 0.19--0.32\tn{46}\tr{h}     & $ 8.0\pm1.6        $\tn{7}  \\
 24 &  IGR J00291       &$\sim2.2\times10^{-12}$\tn{47}\tr{c,s}&  $0.19^{+0.06}_{-0.08}$\tn{2}   &     2.46\tn{48} &   599\tn{48}  &  0.039--0.16\tn{48}         & $ 2.6-3.6          $\tn{1}  \\
 25 &  HETE J1900.1     & $3.9\times10^{-11}$\tn{35}       &  $0.061\pm0.037$\tn{49}             &    1.388\tn{50} &  377\tn{50}   &  0.016--0.07\tn{50}         & $ 4.7\pm0.6        $\tn{7}  \\
 26 &  XTE J1701        &$<9\times10^{-10}$\tn{51}\tr{b}   &  $<5$\tn{51,52}\tr{f}               &      --         &  --           &   --                        & $ 7.3-8.8          $\tn{53}  \\
 27 &  Ter 5 X-2        &$<1.7\times10^{-11}$\tn{54}\tr{f} &  $0.7\pm0.1$\tn{54}                 &  21.274\tn{55}  &  11.04\tn{55} &   $\gtrsim0.4$\tn{55}       & $ 5.5\pm0.9        $\tn{15} \\
 28 &  EXO 0748         & $<4.4\times10^{-10} $\tn{1}      &  $3.8\pm0.2$\tn{56}                 &   3.824\tn{57}  &  552\tn{8}    & $\sim0.1$\tn{58}            & $ 7.4\pm0.9        $\tn{9}  \\
 29 &  1RXS J180408     & $<4.6\times10^{-11} $\tn{59}     &  $0.74_{-0.18}^{+0.09}$\tn{59}\tr{b}&     --          &  --           &   --                        & $ 5.8        $\tn{59}\tr{d} \\
 30 &  Ter 5 X-3        & $<3\times10^{-11} $\tn{60}\tr{f} &  $1.2\pm0.2$\tn{61}\tr{i}           &  --             &  --           &   --                        & $ 5.5\pm0.9        $\tn{15} \\
 31 &  SAX J1750.8      & $2\times10^{-10}$\tn{62}         &  $<2.8$\tn{63}                      &      --         &  601\tn{8}    &   --                        & $ 6.79\!\pm\!0.14  $\tn{9}  \\
 32 &  Swift J1756.9    & $1.5\times10^{-11}  $\tn{35}     &  $<1.0$\tn{38}\tr{b}                &   0.912\tn{64}  &  182\tn{64}   &  0.007--0.03\tn{64}         & $ 8\pm4            $\tn{9}  \\
 33 &  GRS 1747         & $1.0\times10^{-10}  $\tn{65}     &  $<2.4$\tn{65}                      &   12.360\tn{66} &   --          &   --                        & $ 9.5^{+3.3}_{-2.5}$\tn{4} \\
 34 &  IGR J18245       & $\lesssim10^{-10}   $\tn{35}     &  $<0.07$\tn{67}                     &  11.026\tn{68}  &  254\tn{68}   &  $\sim0.2$\tn{68}           & $ 5.5^{+0.2}_{-0.4}$\tn{68,69} \\
 35 &  MAXI J0556       & $<10^{-9}$\tn{70}\tr{f}          &  $<5.1$\tn{71}\tr{f}                &\tn{72}\tr{j}    &   --          &  0.45\tn{72}                & $43.6^{+0.9}_{-1.6}$\tn{71} \\
\noalign{\smallskip}                                                                                                                                                                           
\hline\hline                                                                                      
\end{tabular}                                                                                     
\tablefoot{                                                                                       
\tablefoottext{a}{The statistical errors are evaluated based on the data in the second reference.}
\tablefoottext{b}{Estimated in this work, using data from the given reference.}
\tablefoottext{c}{Donor/accretor mass ratio $M_\mathrm{d}/M_\mathrm{a}=0.18\pm0.06$ [12].}
\tablefoottext{d}{An upper limit and an assumed distance.}
\tablefoottext{e}{Donor/accretor mass ratio $M_\mathrm{d}/M_\mathrm{a}=0.49\pm0.10$ [13].}
\tablefoottext{f}{Conservative upper limit, based on data from the given reference.}
\tablefoottext{g}{Estimated upper bound, including the uncertainty given in this reference.}
\tablefoottext{h}{Donor/accretor mass ratio $M_\mathrm{d}/M_\mathrm{a}=0.123^{+0.012}_{-0.01}$ [34].}
\tablefoottext{i}{The coldest measurement, which agrees with the pre-outburst level.}
\tablefoottext{j}{Two candidate solutions, 16.4~h and 9.75~h.}
\\
\textbf{References:} 
[1]  \citet{CoriatFD12}; 
[2]  \citet{Heinke_ea09};
[3]  \citet{NowakHB02};
[4] \citet{Kuulkers_ea03}; 
[5]  \citet{Ootes_ea16};
[6]  \citet{Merritt_ea16};
[7]  \citet{Watts_ea08};
[8]  \citet{Watts12};
[9]  \citet{Galloway_ea08};
[10] \citet{Heinke_ea07};
[11] \citet{Rutledge_ea99};
[12] \citet{DegenaarWijnands12};
[13] \citet{RiveraSandoval_ea18};
[14] \citet{Matrange_ea17};
[15] \citet{Ortolani_ea07};
[16] \citet{JonkerBW07};
[17] \citet{JonkerNelemans04};
[18] \citet{DegenaarWM13};
[19] \citet{Jonker_ea04};
[20] \citet{Cackett_ea13};
[21] \citet{Ponti_ea18};
[22] \citet{Ortolani_ea99};
[23] \citet{Cackett_ea10_CenX4};
[24] \citet{Cowley_ea88};
[25] \citet{DAvanzo_ea05};
[26] \citet{Jonker_ea07};
[27] \citet{CornelisseWH07};
[28] \citet{TomsickCK07};
[29] \citet{TomsickGK05};
[30] \citet{Casares_ea02};
[31] \citet{Fiocchi_ea09};
[32] \citet{Bilous_ea18};
[33] \citet{Allen_ea18};
[34] \citet{Ootes_ea18};
[35] \citet{VanIH19}, accretion rate is rescaled to the canonical neutron star;
[36] \citet{WalshCB15};
[37] \citet{Sanna_ea16};
[38] \citet{HaskellDH12};
[39] \citet{BultPvdK15};
[40] \citet{Heinke_ea13};
[41] \citet{Galloway_ea02};
[42] \citet{ChakrabartyMorgan98};
[43] \citet{Wang_ea13};
[44] \citet{Riggio_ea08};
[45] \citet{Markwardt_ea02};
[46] \citet{Wang_ea17};
[47] \citet{DeFalco_ea17};
[48] \citet{Galloway_ea05};
[49] \citet{Degenaar_ea17};
[50] \citet{Kaaret_ea06};
[51] \citet{TurlioneAP15};
[52] \citet{Fridriksson_ea11};
[53] \citet{Lin_ea09}
[54] \citet{Ootes_ea19};
[55] \citet{Papitto_ea11};
[56] \citet{Degenaar_ea14_EXO}
[57] \citet{Parmar_ea86};
[58] \citet{MiklesHynes12};
[59] \citet{Parikh_ea18};
[60] \citet{Bahramian_ea14};
[61] \citet{Degenaar_ea15};
[62] \citet{Lowell_ea12};
[63] \citet{ParikhWijnands17};
[64] \citet{Krimm_ea07};
[65] \citet{Vats_ea18};
[66] \citet{intZand_ea03};
[67] \citet{Linares_ea14};
[68] \citet{Papitto_ea13};
[69] \citet{Becker_ea03};
[70] \citet{Homan_ea14};
[71] \citet{Parikh_ea17};
[72] \citet{Cornelisse_ea12}.
}
\end{table*}

It should be noted that the accretion rates are usually evaluated from
observed X-ray luminosities $\tilde{L}_X$ using the equation
\citep[e.g.,][]{VanIH19}
\beq
   \dot{M}_\mathrm{obs} = \frac{\tilde{L}_X R_\mathrm{f}}{
     G M_\mathrm{f}},
\label{Mdotobs}
\eeq
where
$M_\mathrm{f}$ and $R_\mathrm{f}$ are fiducial mass and radius of the
neutron star. In most of the previous works, the ``canonical neutron
star model'' with $M_\mathrm{f}=1.4\,M_\odot$ and $R_\mathrm{f}=10$~km
was used for calculation of $\langle\dot{M}_\mathrm{obs}\rangle$
\citep[e.g.,][]{DegenaarWijnands12}, which is sometimes written  as
$\tilde{L}_X \approx 0.2\,\dot{M}_\mathrm{obs}c^2$. 
\citet{VanIH19} derived
$\dot{M}_\mathrm{obs}$ from $\tilde{L}_X$ assuming
$M_\mathrm{f}=1.4\,M_\odot$ and $R_\mathrm{f}=11.5$~km. Although the
latter radius is more realistic, we have rescaled the
corresponding values of $\langle\dot{M}_\mathrm{obs}\rangle$ in
Table~\ref{tab:obj} (lines 17, 18, 21, 22, 25, 32, and 34) back to the canonical model for
the uniformity of the data sample.

The total (bolometric) accretion luminosity measured at infinity 
is related to the
accretion rate $\dot{M}$ measured locally at the neutron star surface by
equation \citep[e.g.,][]{Mitra98,Meisel_ea18}
\beq
   \tilde{L}_A = \frac{z}{(1+z)^2}\,\dot{M}c^2,
\label{LXvsMdot}
\eeq
where
\beq
   z = (1 - 2GM/Rc^2)^{-1/2} - 1
\label{z}
\eeq
is the gravitational redshift at the stellar surface and
$\tilde{L}_A=A\tilde{L}_X$, $A>1$ being the bolometric correction. 
Excluding
$\tilde{L}_X$ from Eqs.~(\ref{Mdotobs}) and (\ref{LXvsMdot}), 
we obtain
\beq
   \dot{M}_\mathrm{obs} = \frac{z}{z_\mathrm{f}}
      \left(\frac{1+z_\mathrm{f}}{1+z} \right)^2\,
      \frac{\dot{M}/A}{1+z_\mathrm{f}/2},
\label{MdotobsMdot}
\eeq
where $z_\mathrm{f}$ is the fiducial gravitational redshift, given
by \req{z} with $M=M_\mathrm{f}$ and $R=R_\mathrm{f}$
($z_\mathrm{f}=0.3057$ for the canonical neutron star model).

Most of the values in Table~\ref{tab:obj} are taken from papers
indicated by the numbers in square brackets. The cases where the listed
values are not plainly adopted, but derived in this work from the given
references, are marked by a footnote to the table. In particular,
whenever
the uncertainties of bolometric luminosities are not given explicitly, 
we evaluate them from the uncertainties of effective temperatures and/or
the scattering of results obtained with using different spectral models.
For sources with just one observed outburst, the average mass accretion
rate is reported in Table \ref{tab:obj} as an  upper limit, because the
duration of the quiescent period can be much larger than X-ray
observation time-line ($\sim 30$~yr).%
\footnote{According to disk instability model, maximal duration of the
quiescence period $\sim 180$~yr (see, e.g., section 6.4 in
\citealt{Lasota01}). Estimates by \citet{ChugunovGK14} suggest that
even longer quiescence period ($\sim 1000$~yr) should be allowed, if
all X-ray sources known as candidate quiescent LMXBs in globular
clusters, are indeed LMXBs in quiescent state. These candidate sources
are selected by X-ray spectrum, which is well fitted by neutron star
thermal emission, but they are treated as candidates because no outburst
from these source have been detected yet (see, e.g., \citealt{Bahramian_ea15}
and references therein). An alternative explanation of these sources,
suggested by \citet{ChugunovGK14}, is based on heating associated with
Chandrasekhar-Friedman-Schutz instability \citep{FS78a,FS78b} and does
not require such a long quiescence time in LMXBs.}
Note that in many cases there can be much larger systematic
uncertainties due to unaccounted model-dependence,  poorly known
distance or hydrogen column density, etc., therefore the listed errors
should be considered as lower limits to largely unknown actual
uncertainties.

Some of the considered SXTs or listed numbers
deserve the following additional comments. 
\begin{itemize}

\item\emph{1. 4U 2129+47}. The bolometric luminosity is derived from
the effective temperature and radius, obtained by \citet{NowakHB02} for
the canonical neutron star model (spectral fit models E, F in table~2
of that paper), and its uncertainties are roughly estimated from the
given temperature uncertainties and scattering of different estimates
in that reference. The same value of $\tilde{L}_\mathrm{q}$ has been
given by \citet{Heinke_ea09}.

\item\emph{2. KS1731$-$260}. The average accretion rate is estimated as
the average rate during the outburst ($1.5\times10^{17}$ g s$^{-1}$,
\citealt{Ootes_ea16}) multiplied by the observed outburst duration (11.5
years) and divided by the total time of observations ($\approx 30$
years). This estimate is very uncertain, because only one transition
between the active and quiescent states (the end of the outburst in
2001) has been observed. We derive the statistical errors  on
$\tilde{L}_\mathrm{q}$ from the errors on the effective
temperature given by \citet{Merritt_ea16}, however the authors warn that
there can be large systematic errors. 

\item\emph{4. EXO 1745$-$248 (Ter 5 X-1)}. The luminosity  is
completely dominated by non-thermal emission
\citep[e.g.,][]{DegenaarWijnands12}. The limit
$\tilde{L}_\mathrm{q}<2.1\times10^{33}$ erg~s$^{-1}$ was most often
quoted \citep[e.g.,][]{Heinke_ea07,BeznogovYakovlev15a}.
\citet{RiveraSandoval_ea18} found a strong variability of X-ray
luminosity of this SXT in quiescence, with a luminosity variation in the
0.5\,--\,10 keV energy range from $3\times10^{31}$ erg~s$^{-1}$ to
$2\times10^{34}$ erg~s$^{-1}$. Since the total, thermal and non-thermal
luminosity cannot be smaller than the quasi-equilibrium thermal
component, we adopt $3\times10^{31}$ erg~s$^{-1}$ as an upper limit to
this component in the X-rays. For the bolometric quiescent luminosity,
this implies the conservative upper limit $\tilde{L}_\mathrm{q}<10^{32}$
erg~s$^{-1}$, which is consistent, within uncertainties, with the
constraint $\tilde{L}_\mathrm{q}\lesssim7\times10^{31}$ erg~s$^{-1}$,
obtained by \citet{DegenaarWijnands12} in frames of a specific spectral
model. 

\item\emph{7. MXB 1659$-$29}. This bolometric luminosity corresponds to
the effective temperature estimate $k\tilde{T}_\mathrm{eff}=55\pm3$ eV,
obtained by \citet{Cackett_ea13} for observations performed 11 yr after
the end of the outburst. This estimate is consistent with the previous
two, for observations taken approximately 4 and 7 years earlier
\citep{Cackett_ea06,Cackett_ea08}. However, the count rate has dropped
by a factor of 3 in the latest observation compared with the previous
two (possibly due to an increase in hydrogen column density, see
discussion in \citealt{Cackett_ea13}). Inclusion of a power-law
component  improves the spectral fit and gives
$k\tilde{T}_\mathrm{eff}\approx45$ eV 
($\tilde{L}_\mathrm{q}\sim 9 \times10^{31}$~erg\,s$^{-1}$),
which is reflected in the larger negative error
estimate in our table, but the power-law component is not required to
fit the previous observations \citep{Cackett_ea13}.

\item\emph{9. Cen X-4.} Luminosity estimate is based on the temperature 
reported in table~2 of \citet{Cackett_ea10_CenX4} for Suzaku observation 
(the latest and the coldest one) 
and the canonical neutron star parameters.

\item\emph{16. 4U 1908+005 (Aql X-1)}. The long-term average accretion
rate is calculated as the total accreted mass during the period of
regular observations from 1996 to 2015, determined from table~2 of
\citet{Ootes_ea18}, divided by this time interval.  We note that this
rate was  $\sim40\%$ higher in the first five years of this period.
Because of frequent outbursts, this neutron star never reaches thermal
equilibrium \citep{Ootes_ea18}. Here, the baseline quiescent luminosity at infinity
$\tilde{L}_\mathrm{q}$ is calculated from the range of base levels of
the effective temperature in the numerical fitting simulations by
\citet{Ootes_ea18}, for the adopted values of $M=1.6\,M_\odot$ and
$R=11$~km.

\item\emph{19. XTE J0929$-$314.} This limit on
$\langle\dot{M}_\mathrm{obs}\rangle$ roughly agrees with the refined
constraint
$\langle\dot{M}_\mathrm{obs}\rangle<(8.4^{+22}_{-6.7})\times10^{-12}\,M_\odot/$yr
(\citealt{Heinke_ea13,VanIH19}; quoted value is scaled to canonical
neutron star).

\item\emph{20. SAX J1808.4$-$3658}. From different estimates of the
upper limit on $\tilde{L}_\mathrm{q}$ given by \citet{Heinke_ea09}
($4.9\times10^{30}$ erg s$^{-1}$, $6.2\times10^{30}$ erg s$^{-1}$,  and
$1.3^{+0.6}_{-0.8}\times10^{31}$ erg s$^{-1}$), corresponding to
different spectral models applied to fit the continuum, we have selected
the highest estimate as the most conservative option.

\item\emph{21. XTE J1807$-$294}. More precisely,
$\langle\dot{M}_\mathrm{obs}\rangle<(1.3^{+1.7}_{-1.0})\times10^{-11}\,M_\odot/$yr
(\citealt{VanIH19}; here it is scaled to the canonical neutron star
model).

\item\emph{24. IGR J00291+5934.} The given value of accretion rate  is
calculated using \req{Mdotobs} by summation of the fluences listed in
table 2 of \citet{DeFalco_ea17} for four last outbursts, divided by the
time interval covering these outbursts and preceding periods of
quiescence  ($\Delta t = 13.8$ yr from September 2001 to July 2015),
for the distance $4.2$~kpc derived by these authors. This estimate well
agrees with $\langle\dot{M}_\mathrm{obs}\rangle
\approx2.5\times10^{-12}\,M_\odot$ yr$^{-1}$ in table~2 of
\citet{Heinke_ea09},  based on three outbursts. It is approximately
twice as large as the rate  reported by \citet{VanIH19} (presumably due
to the lower distance estimate, 2.6\,--\,3.6 kpc, used in that paper)
and approximately two times smaller than the rate reported by
\citet{CoriatFD12}  (likely due to small interval of time averaging,
$\Delta t\sim3$~yr). A substantial part of the quiescent emission is
non-thermal, perhaps due to a residual accretion disk
\citep{Torres_ea08,Baglio_ea17}.

\item\emph{25. HETE J1900.1$-$2455}. The quoted estimate of
$\tilde{L}_\mathrm{q}$ is based on an analysis of several non-detections
and a single detection of this source in quiescence, carried out by
\citet{Degenaar_ea17}, who have shown that the crust may have not fully
relaxed by the time of this detection (the likely quiescent base
luminosity values are accommodated by the quoted uncertainties).

\item\emph{26. XTE J1701$-$462}. We estimate
$\langle\dot{M}_\mathrm{obs}\rangle$ by multiplying outburst  $\dot{M}$ from
table~1 of \citet{TurlioneAP15} by the outburst duration (1.6 years) and
dividing by the fiducial time-line of  X-ray observations (30 years). Since
\citet{TurlioneAP15} noted that this source may not have reached
equilibrium, we take the smallest observed luminosity
as an upper bound on $\tilde{L}_\mathrm{q}$.

\item\emph{27. IGR J17480$-$2446 (Ter 5 X-2)}.
$\langle\dot{M}_\mathrm{obs}\rangle$ is estimated from the outburst
level $\langle\dot{M}\rangle=3\times10^{-9}$ (11\% of the Eddington
limit) times the rough estimate of the duty cycle: 2 months of outburst
in 2010 over observation timescale of 30 yr. Note that the spin
frequency of this neutron star is relatively small ($11$~Hz,
\citealt{Papitto_ea11}), suggesting that the total accreted mass is
probably small and the crust may be not fully replaced yet by accreted
material \citep{WijnandsDP13}.

\item \emph{28. EXO 0748$-$676}. The quoted $\tilde{L}_\mathrm{q}$
corresponds to the last observation reported by
\citet{Degenaar_ea14_EXO}. It should be noted that  the same reference
reports pre-outburst detection in 1980 by Einstein observatory with
$\tilde{L}=2.3\pm 1.2$~erg\,s$^{-1}$, which is compatible with the
quoted estimate at the $\approx1.3\sigma$ level.

\item \emph{29. 1RXS J180408.9$-$342058}. Luminosity is estimated on the
base of temperature confidence interval in table~2 of
\citet{Parikh_ea18} for observation 3 (2.4 yr since the end of
outburst) and fit with unfixed power-law index, for the assumed neutron
star mass $M=1.6\,M_\odot$ and radius $R=11$~km. It agrees with the
X-ray luminosity value in this table with bolometric correction.

\item\emph{30. Swift J174805.3$-$244637 (Ter 5 X-3)}. The estimate of 
$\langle\dot{M}_\mathrm{obs}\rangle$ \citep{Bahramian_ea14} includes an
outburst of 2002, which is not firmly attributed to this source;
otherwise it could be smaller. For $\tilde{L}_\mathrm{q}$, we take the
coldest measurement from \citet{Degenaar_ea15},  which agrees with the
pre-outburst level.

\item\emph{33. Swift J1750.7$-$3117 (GRS 1747$-$312)}. According to
\citet{Vats_ea18}, about of a half of observed flux is thermal; perhaps
there is a residual accretion.

\item\emph{34  IGR J18245-2452}. This object is known to switch between
accretion and rotation powered pulsar states (transitional millisecond
pulsar, \citealt{Papitto_ea13}). 

\item\emph{35. MAXI J0556$-$332}. $\langle\dot{M}_\mathrm{obs}\rangle$
is estimated as Eddington-limited accretion during 480 days of outburst
\citep{Homan_ea14} averaged over 30 years of X-ray observation time-line. For the
luminosity, we take the minimal value from several observations of
\citet{Parikh_ea17} and treat it as the upper bound, because the
quasi-equilibrium state may have not yet been reached.

\end{itemize}

\begin{figure}
\centering
\includegraphics[width=\columnwidth]{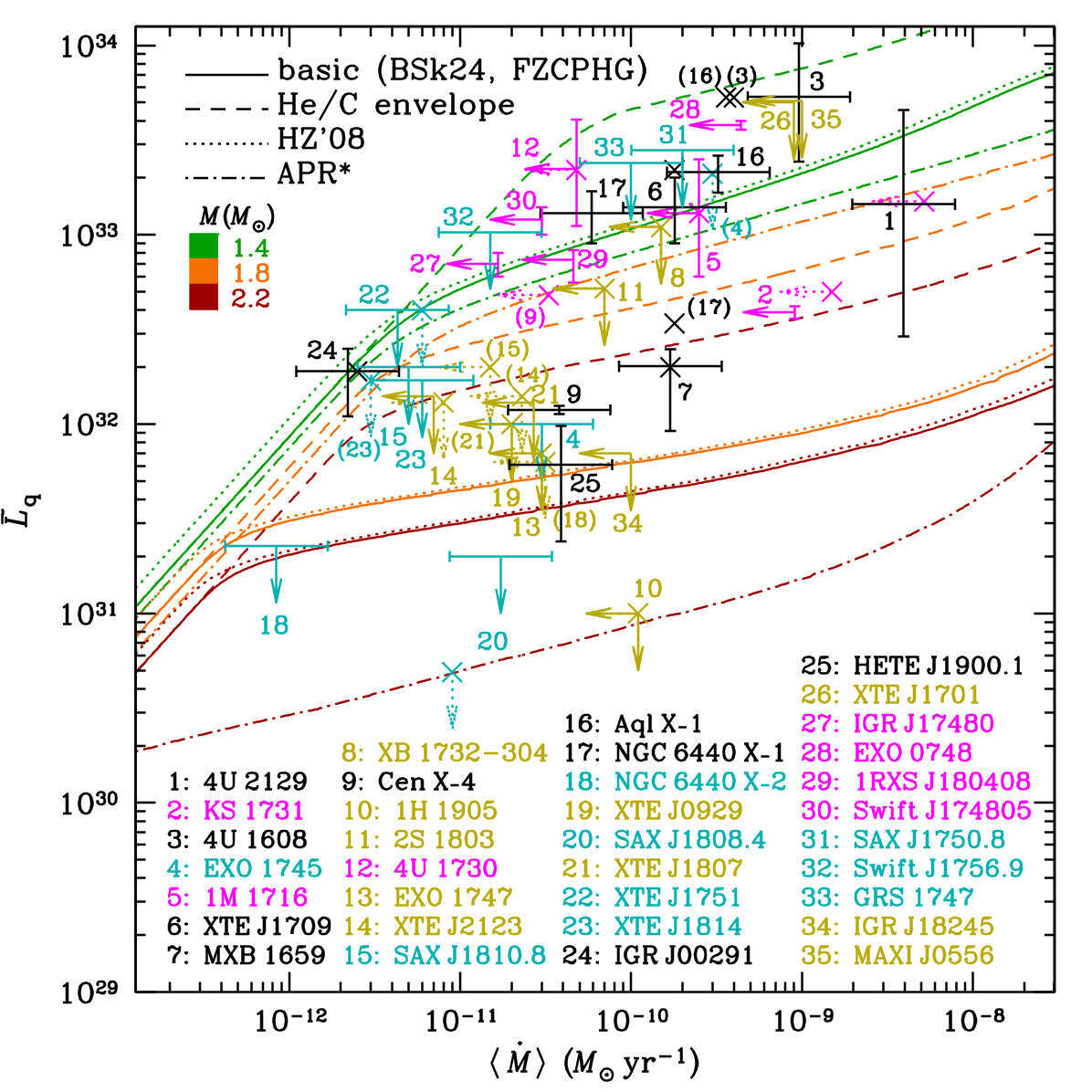}
\caption{Quiescent thermal luminosities of SXTs as functions of  average
accretion rates. Solid errorbars and arrows show the data listed in
Table~\ref{tab:obj}. Crosses without errorbars and dotted arrows show
the older estimates and upper limits (e.g., table~2 of
\citealt{BeznogovYakovlev15a}). The solid errorbars (arrows) are labeled
by numbers from the first column of Tables~\ref{tab:objlist}
and~\ref{tab:obj}, and abbreviated names of associated objects are
listed in the lower part of the figure. In some cases, to avoid
confusion, the crosses or dotted  arrows are also labeled by the
numbers in parentheses. The lines show theoretical predictions for the
thermal quasi-equilibrium luminosity as a function of time-averaged
accretion rate due to the heating of a fully accreted crust of neutron
stars of three masses, $M=1.4\,M_\odot$, $1.8\,M_\odot$, and
$2.2\,M_\odot$ (from upper to lower lines of the same type, coded by
color), computed using the method described in Sect.~\ref{sect:YLH} for
different theoretical models described in Sect.~\ref{sect:physics}: the
``basic model'' (solid lines), the same model with a fully accreted
heat-blanketing envelope (dashed lines), the basic model with HZ'08
heating and composition of the crust instead of FZCPHG (the dotted
lines), and the alternative model (APR$^*$ EoS and composition, BBST
effective baryon masses) with iron thermal-insulating envelope
(dot-dashed lines).
}
\label{fig:compdat}
\end{figure}

Figure~\ref{fig:compdat} shows the redshifted thermal quasi-equilibrium
luminosities of the neutron stars in the SXTs in quiescence,
$\tilde{L}_\mathrm{q}$, and their average accretion rates,
$\langle\dot{M}_\mathrm{obs}\rangle$, inferred from observations. The
estimates of $\tilde{L}_\mathrm{q}$ and
$\langle\dot{M}_\mathrm{obs}\rangle$ are plotted in
Fig.~\ref{fig:compdat} by errorbars, and the upper bounds are indicated
by arrows. The errors are provisionally set to a factor of two in the
average accretion rates. Such errorbars appear to approximately
represent the anticipated magnitude of cumulative statistical and
systematic errors. For the luminosities, we use  errors from
Table~\ref{tab:obj}, which do not include possible  systematic
uncertainties (except when especially noted). The most important sources
of errors in most cases appear to be, for $\langle\dot{M}\rangle$, the
lack of reliable observations on a long timeline and in some cases  the
uncertainty in the distance,%
\footnote{
As recently shown by \citet{CarboneWijnands19}, a bias can be associated
with estimation of the duty cycle
 (that is the fraction of time in the active state): missing of the outbursts decreases
estimated duty cycle for  systems with rare outburst, while  variability
of the duty cycle can
lead to overestimation of the duty cycle for systems with frequent
outbursts.
}
and for $L_\mathrm{q}$, the uncertainty in
the distance, to which in some cases are added uncertainties in spectral
decomposition and emission models. For comparison, along with the data
from Table~\ref{tab:obj} we plot the traditional dataset, which has been
used for similar illustrations up to now
\citep{Heinke_ea10,WijnandsDP13,WijnandsDP17,BeznogovYakovlev15a,BeznogovYakovlev15b,Fortin_ea18}.
In the cases where both $\langle\dot{M}\rangle$ and $L_\mathrm{q}$ have
been estimated, the errorbars are plotted in black, while different
colors are chosen to show the cases where one of these quantities or
both of them have only upper limits. In the bottom of the figure,
truncated names of the SXT sources are listed for easy reference. The
lines in Fig.~\ref{fig:compdat} show theoretical functions
$\tilde{L}(\langle\dot{M}\rangle)$ explained in Sect.~\ref{sect:YLH}. 

\subsection{Simple evaluation of quiescent luminosity}
\label{sect:YLH}

The timescale of thermal relaxation of neutron-star crust is much
shorter than the neutron-star cooling timescale (see, e.g.,
\citealt{GYP01}). Therefore, after the accretion halts, the neutron star
relaxes to thermal quasi-equilibrium, which is determined by neglecting
the slow variations of the thermal state of the stellar core
(cf.~\citealt{Colpi_ea01}). The quasi-equilibrium temperature
distribution is controlled by the redshifted temperature of the core
$\tilde{T}_\mathrm{core}$, which is nearly constant because of the high
thermal conductivity in the core. Therefore, a quiescent state of a
neutron star in an SXT should be the same as the state of a cooling
isolated neutron star (INS) with the same mass and composition at the
age when this virtual INS would have the same $\tilde{T}_\mathrm{core}$
as the considered neutron star in the SXT at quiescence. 

This similarity is often used to determine  quasi-equilibrium quiescent
thermal luminosities of the SXTs, following the method suggested by
\citet{YakovlevLH03}.  This method assumes that the \emph{total} energy
loss by a neutron star in the quasi-equilibrium state
equals the heat deposited by the deep crustal heating
over a period covering many cycles of outbursts and
quiescence. This assumption can be written as
$\tilde{L}_\mathrm{tot}=\langle\tilde{L}_\mathrm{h}\rangle$, where
\beq
   \tilde{L}_\mathrm{tot}=\tilde{L}_\nu+\tilde{L}_\gamma
\label{Ltot}
\eeq
 is the total energy loss in unit time as measured by a distant
observer, $\tilde{L}_\mathrm{h}$ and $\tilde{L}_\nu$ are the redshifted
heating power and neutrino luminosity given by
\req{tL}, and $\tilde{L}_\gamma$ is the measured bolometric photon
luminosity of the INS, which is assumed to be equal to the
quasi-equilibrium measured bolometric photon
luminosity of the SXT in quiescence, $\tilde{L}_\mathrm{q}$.
 
\begin{figure}
\centering
\includegraphics[width=\columnwidth]{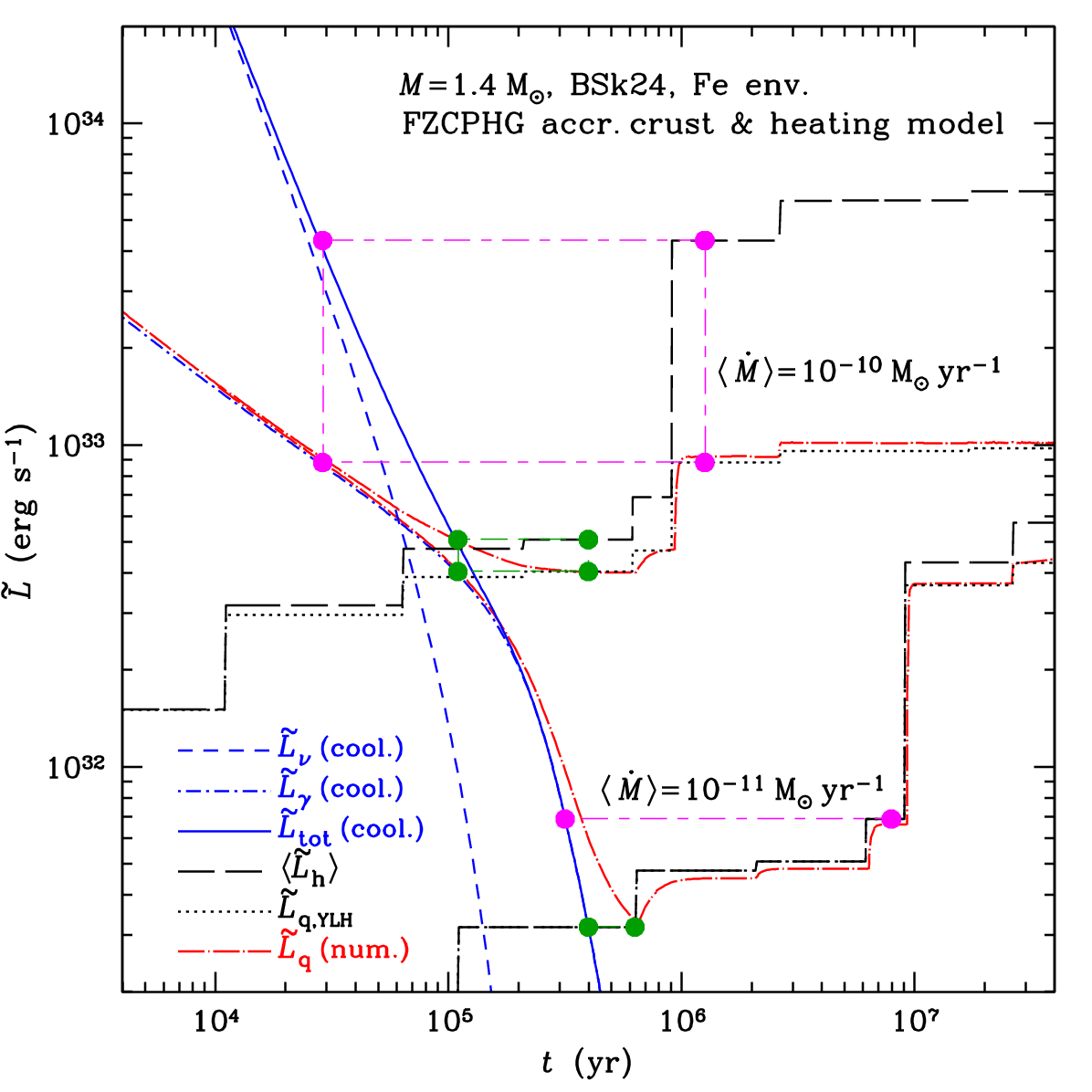}
\caption{Illustration of the approximate calculation of the
quasi-equilibrium quiescent luminosity according to the method of
\citet{YakovlevLH03} (YLH) for a neutron star with mass
$M=1.4\,M_\odot$, described by the BSk24 EoS and composition model in
the nonaccreted crust and the core and by the FZCPHG model of
composition and heat sources in the accreted crust, with an iron
heat-blanketing envelope. The blue solid line shows the total energy
loss rate $\tilde{L}_\mathrm{tot}$, which is the sum of the neutrino
luminosity $L_\nu$ (blue short-dashed line) and the photon luminosity
$L_\gamma$ (blue dot-short-dash line), of a cooling INS as function of
the cooling time $t$. The black long-dashed stepped lines show the
average power $\langle\tilde{L}_\mathrm{h}\rangle$ as function of the
accretion time $t$, assuming a constant average accretion rate
$\langle\dot{M}\rangle=10^{-10}\,M_\odot/$yr (the upper line) or
$\langle\dot{M}\rangle=10^{-11}\,M_\odot/$yr (the lower line). The
dotted stepped lines show the photon luminosities $\tilde{L}_\gamma$, as
functions of the accretion time, that correspond to the cooling time
moments when
$\tilde{L}_\mathrm{tot}=\langle\tilde{L}_\mathrm{h}\rangle$. The thin
long-dash--short-dash lines serve as guides to eye: they connect the
corresponding total and photon luminosities and the corresponding
cooling and heating time values. The red dot-long-dashed lines show the
evolution of the bolometric photon luminosity in the numerical model of
a cooling and heating neutron star, assuming that the accretion starts
sufficiently soon after the start of the cooling and proceeds at a rate
either $10^{-10}\,M_\odot/$yr (the upper curve) or
$10^{-11}\,M_\odot/$yr (the lower line). All
plotted luminosities are redshifted as measured in a remote frame of
reference.
}
\label{fig:nsgtL}
\end{figure}

This method is illustrated in Fig.~\ref{fig:nsgtL}. Here, the average
heating power is calculated according to the model FZCPHG, using
\req{tL}, which in the case of heat sources concentrated at a series of
thin shells turns into
\beq
   \tilde{L}_\mathrm{h} = \sum_{i}
     \ee^{2\Phi_i/c^2} W_i,
\label{tLh}
\eeq
where $i$ enumerates the reaction shells in the order of increasing
pressure, $r_i$ is the radius at the given shell, $M_{r,i}$, $\Phi_i$,
and $W_i$ are the respective values of $M_r$, $\Phi(r)$, and 
heating power generated at the given surface. The last quantity is
given by the relation
\bea
   W_i=\dot{a}\cdot E_{\mathrm{h},i}=
   3.8\times10^{49}\,(\dot{M}/M_\odot\mbox{~yr}^{-1})
   \, E_{\mathrm{h},i}\mbox{~s}^{-1}
   \\
   \approx 6\times10^{34}\,\dot{M}_{-9} \,(E_{\mathrm{h},i} /\mbox{MeV})
   \mbox{~erg~s}^{-1},
\label{powerheat}
\eea
 where $\dot{a}$ is the number of accreted baryons per unit time,
$E_{\mathrm{h},i}$ is the released energy per baryon at the $i$th
reaction shell, and $\dot{M}_{-9}$ is the
accretion rate in units of $10^{-9}\,M_\odot/$yr in the local reference
frame. The summation in \req{tLh} is performed only for those shells
that lie within the accreted part of the crust, which means that the
total mass above a given shell, $\Delta M= M-M_{r,i}$, is smaller than the
total accreted mass $\Delta M_\mathrm{acc}$.
Here and hereafter, following the previous works, we neglect the heat
that is released due to compression of the pristine ground-state 
crust in the course of accretion, 
assuming it to be smaller than the heat produced by the nuclear
reactions in the accreted crust.

The lines in Fig.~\ref{fig:compdat} show theoretical redshifted thermal
quasi-equilibrium luminosities in quiescence, $\tilde{L}_\mathrm{q}$, as
functions of the average accretion rate. The luminosities have been
computed under the assumption of fully accreted crust, which provides
the maximum deep heating power by including all the reaction shells in
the sum in \req{tLh}. The physics input is described in
Sect.~\ref{sect:physics}. Our basic model includes the BSk24 EoS and
composition of the non-accreted part of the star, FZCPHG model of
heating and composition of the accreted crust, the MSH, BS, and BEEHS
models for different types of superfluidity, and the iron
heat-blanketing envelope. The alternative model, labeled APR$^*$,
employs the APR$^*$ EoS and composition of the core and the BBST
effective masses of the nucleons. Additional modifications include
allowance for the accreted heat-blanketing envelopes composed of helium
and carbon instead of iron (see, e.g., \citealt{PCY97,BeznogovPY16}) or
the use of the HZ'08 model for the composition and heating of the
accreted crust. The accretion rates in the local reference frame,
$\dot{M}$, have been used as input for obtaining the heating power
density $Q_\mathrm{h}$, but in the figure they have been converted into
$\dot{M}_\mathrm{obs}$ according to \req{MdotobsMdot} (with $A=1$) for
direct comparison with the data listed in Table~\ref{tab:obj}.

A comparison of the theoretical heating curves and the observational
data in Fig.~\ref{fig:compdat} shows that the quiescent luminosities of
the hottest sources in the upper part of the figure can only be
explained if we suppose that they have accreted heat blanketing
envelopes, whereas one of the coldest SXTs, SAX J1808.4$-$3658, requires
a very massive neutron star for its explanation and is not compatible 
with some theoretical models of the neutron star matter: in our case, it
can be described by the APR$^*$ model of a neutron star with mass
$M=2.2\,M_\odot$ and iron envelope, but not by the BSk24 model. This
difference is related to the larger stiffness of the BSk24 EoS, which
leads to smaller central densities of the most massive neutron star
models and consequently to lower intensities of the direct Urca process,
compared to the APR$^*$ models.

In Fig.~\ref{fig:nsgtL}, various neutron-star luminosities are plotted
as functions of time $t$, assuming the simplified model, in which
accretion proceeds at a constant average rate
$\langle\dot{M}\rangle$ and starts sufficiently soon after the start of the cooling that
the difference between the cooling age and accretion duration can be
neglected.
In this case $\Delta
M_\mathrm{acc}=\langle\dot{M}\rangle\,t$.  Most of the time-dependencies
shown below imply these minimal assumptions. The lines corresponding to
two fixed values of the average accretion rate are shown,
$\langle\dot{M}\rangle=10^{-10}\,M_\odot/$yr and
$\langle\dot{M}\rangle=10^{-11}\,M_\odot/$yr. 

For each accretion rate, one of the lines shows the average redshifted
heating power $\tilde{L}_\mathrm{h}$. The  ``steps'' on  this line
correspond to the moments $t$, when the accreted matter starts to
involve a new reaction shell, so that a new discrete heat source is
included in the sum (\ref{tLh}); between these moments
$\tilde{L}_\mathrm{h}$ is constant, so the line is horizontal. 

In the same figure, we have plotted $\tilde{L}_\nu$, $\tilde{L}_\gamma$,
and $\tilde{L}_\mathrm{tot}$ [\req{Ltot}]
for a cooling neutron star as functions of
the cooling time and the photon luminosity in quiescence
$\tilde{L}_\mathrm{q}$, calculated
according to the \citet{YakovlevLH03} (YLH) method. 
In this case, the quasi-equilibrium luminosity in quiescence
$\tilde{L}_\mathrm{q}$ increases in steps as a function of the accretion
time, following the steps of $\tilde{L}_\mathrm{h}$. 
The maximum $\tilde{L}_\mathrm{q}$ value is reached when the
innermost reaction shell has become included in the accreted crust.  In
the FZCPHG model this occurs when the accreted matter is pushed to
density of the second-last reaction shell
$\rho=1.7\times10^{13}$~\gcc{}. Pushing it further to the last shell at
$\rho=7.3\times10^{13}$~\gcc{}  does not increase $L_\mathrm{h}$,
because the heating power $W$ at the last shell is negligible. For the
neutron star model in Fig.~\ref{fig:nsgtL} ($M=1.4\,M_\odot$, the BSk24
and FZCPHG models for the nonaccreted and accreted matter, respectively),
the saturation of the heating power occurs at accretion time
$t\approx1.7\times10^{-3}\,(\langle\dot{M}\rangle/M_\odot)^{-1}$.
At earlier epochs, the heating power and the respective quiescent
luminosity are smaller.

\begin{figure}
\centering
\includegraphics[width=\columnwidth]{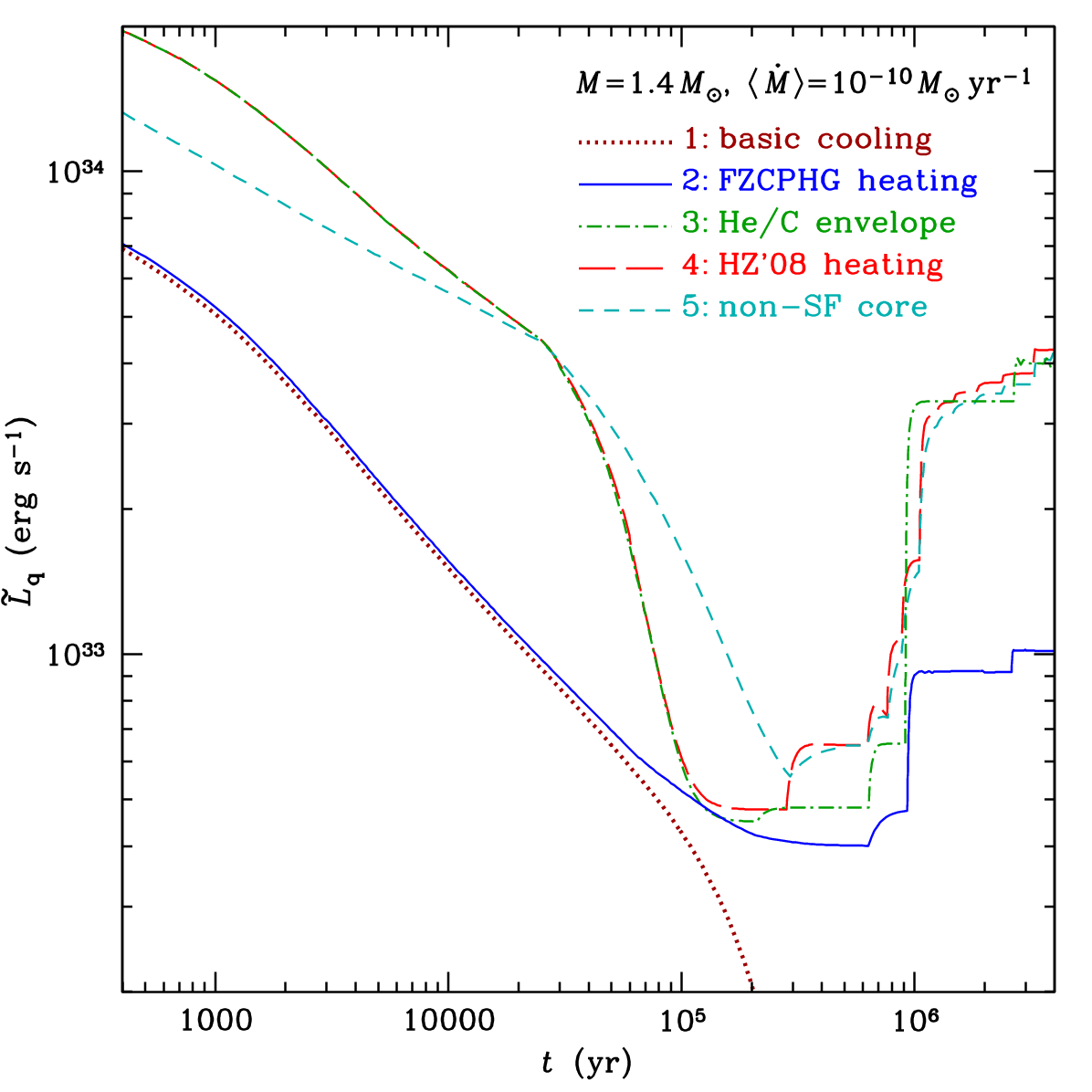}
\caption{Illustration of the effects of differences in the physics input
on the long-term evolution of a neutron star.
Redshifted thermal
luminosity is shown as function of time. Dotted line (1) -- the basic
cooling model; solid line (2) -- the same model with heating according to 
the FZCPHG model, calculated assuming that the accretion starts shortly
after the birth of the neutron star with constant average rate of
$10^{-10}\,M_\odot$/yr; dot-dashed line (3)  -- the same with replacement of
the iron heat-insulating envelope to an accreted He/C envelope,
long-dashed line (4) -- the same accreted envelope, but the alternative
heating model, short-dashed line (5) -- 
the same but without baryon superfluidity in the core.
}
\label{fig:nsgtvar}
\end{figure}

\begin{figure}
\centering
\includegraphics[width=\columnwidth]{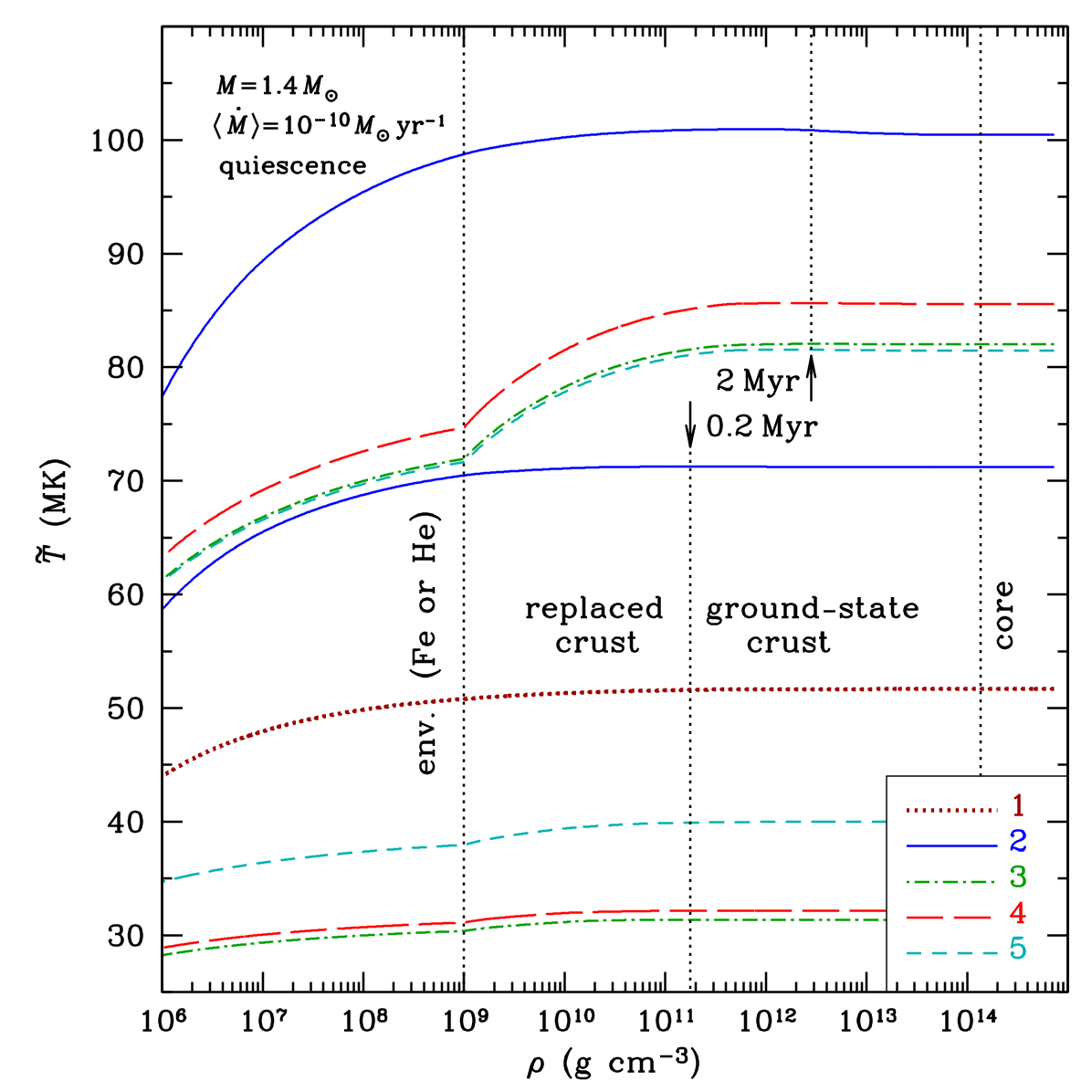}
\caption{Redshifted temperature profiles for the same models
as in Fig.~\ref{fig:nsgtvar}
drawn with the same line styles at $t=2\times10^5$~yr (the lower
curves) and $t=2\times10^6$~yr (the upper curves). The vertical dotted
lines separate the iron or accreted envelope, the accreted (replaced) crust,
the non-accreted (ground-state) crust, and the core.
}
\label{fig:nsgrvar}
\end{figure}

\section{Long-term thermal evolution}

We have compared the YLH method with the results of our accurate
numerical simulations of the evolution of a cooling and heating neutron
star. Figure~\ref{fig:nsgtL}
presents the bolometric photon luminosity as function of time.  In the
numerical model, the photon luminosity decreases at early age, when it
is dominated by the heat initially stored in the interior of the neutron
star and the crust has mainly ground-state composition, so that the deep
crustal heating is negligible. The luminosity has a minimum at an
intermediate age $\sim10^5-10^6$~yr (depending on
$\langle\dot{M}\rangle$) and then increases due to the increasing
thickness of the accreted part of the crust. 
We assume that the crust is
initially ground-state, but gradually it is being replaced by the
accreted crust. The boundary between the accreted crust and the
ground-state crust is determined by the accreted mass $\Delta
M_\mathrm{acc} = \langle\dot{M}\rangle\,t$.
When the reprocessed
accreted matter reaches a new reaction shell, $\tilde{L}_\mathrm{q}$
starts to increase, first sharply and then slowly approaching the new
quasi-stationary value. The comparison shows that the YLH model
accurately predicts the quasi-stationary values of the redshifted
bolometric luminosity in quiescence, $\tilde{L}_\mathrm{q}$, although it
does not reproduce details of transitions from one  quasi-stationary
value to the next one. In reality, $\tilde{L}_\mathrm{q}$ follows
$\tilde{L}_\mathrm{h}$ not immediately. Instead, it gradually approaches
the equilibrium values predicted by the YLH model. With increasing
accretion time, this delay becomes less and less significant 
(in comparison with the age of the star),
so that
for the old SXTs the YLH method proves to be very accurate. We also see
that the minimum value $\tilde{L}_\mathrm{q,min}$ is rather accurately
determined by the intersection of the stepped line representing the YLH
model and the INS cooling curve.

In Fig.~\ref{fig:nsgtvar} we examine the effects of
several alterations in 
the models of outer envelopes, crust, and core, and two different
heating models on the long-term average evolution of the thermal
luminosity of an accreting neutron star of the ``canonical'' mass
$M=1.4\,M_\odot$. Here, the thermal evolution computed
using the same model as in Fig.~\ref{fig:nsgtL} is compared to the
analogous computations but with replacement of some ingredients of
the theoretical model to their alternatives. 

First, we include an accreted envelope instead of the standard iron
envelope, which extends to $\rho=10^9$ \gcc. The light-element
blanketing envelope is more transparent to heat, therefore the surface
luminosity is higher.  Next, keeping the accreted envelope unchanged, we
replace the FZCPHG accreted-crust model by the HZ'08 model. The effect
of this replacement is sensible, although less dramatic than the effect
of the accreted outer envelope. Then we explore the effects of the
superfluidity and the effective baryon masses. A change of the
superfluidity model in the crust from MSH to GIPSF and a change of the
baryon effective mass model almost do not affect the thermal evolution,
so that the corresponding curves would be practically indistinguishable
if plotted in Fig.~\ref{fig:nsgtvar}. At contrast, switching-off
superfluidity in the core is seen to have a substantial effect,
particularly at the cooling stage. 

Figure \ref{fig:nsgrvar} shows the corresponding internal temperature
profiles at two moments of time, 200 kyr and 2 Myr. One can see the
breaks on the profiles 3\,--\,5 at density $\rho=10^9$ \gcc, which
limits the helium accreted crust in these three models: the smaller
slopes of the lines reflect the lower thermal conductivity for lighter
chemical elements. In this figure we also mark the boundaries of the
replaced accreted crust layer, where the heating sources are confined.
The thickness of this layer is larger for the larger accretion duration
(2 Myr) than for smaller one (200 kyr), therefore it includes more
heating sources (cf.{} Fig.~\ref{fig:heatsrc}), which explains the
higher positions of the temperature profiles for $t=2$~Myr. The higher
position of the profile 2, calculated assuming the iron heat-blanketing
envelope, is explained by the better thermal insulation provided by this
envelope. The same insulation results in cooler surface layers (beyond
the figure frame) and the lower photon flux seen in
Fig.~\ref{fig:nsgtvar}.

\begin{figure}
\centering
\includegraphics[width=\columnwidth]{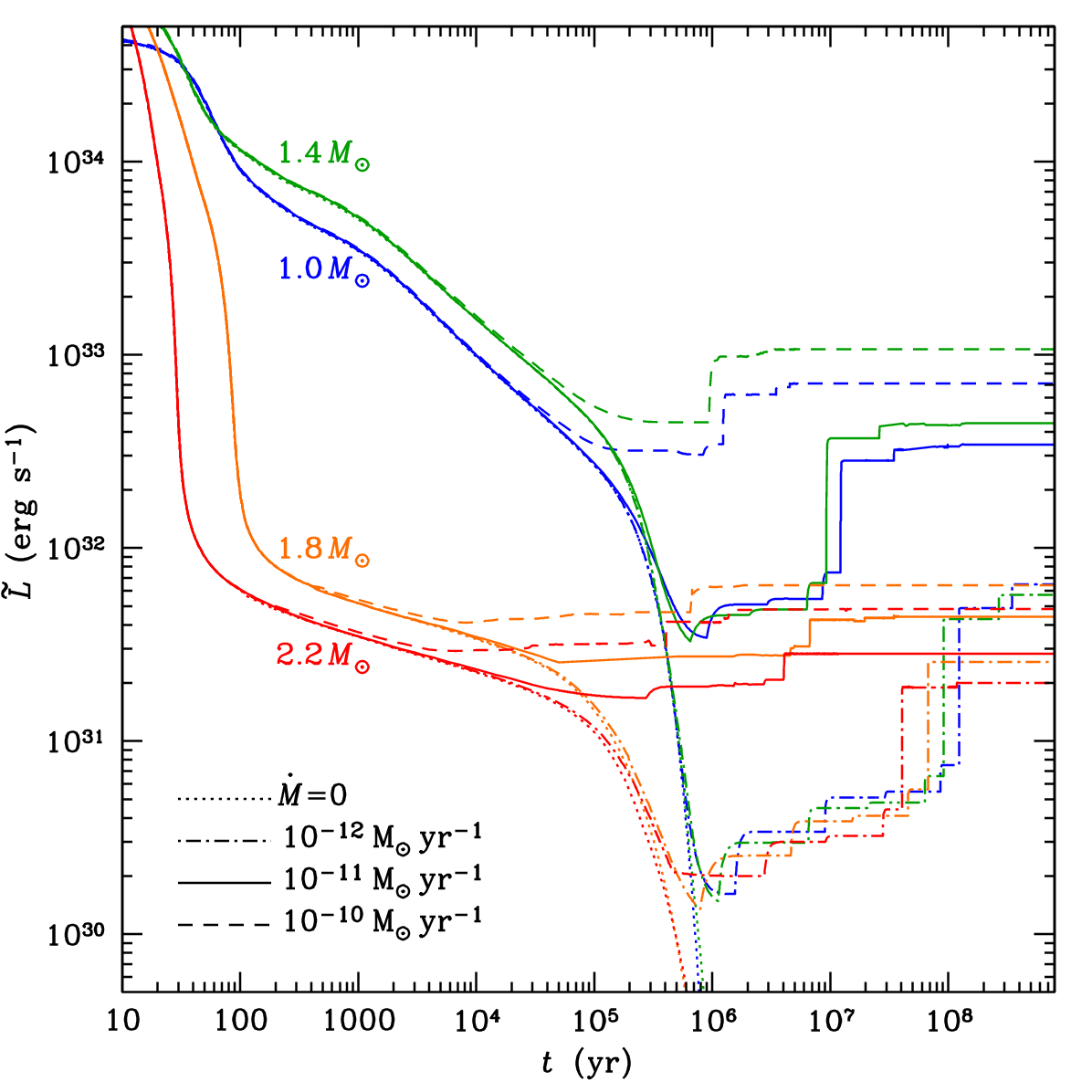}
\caption{Examples of the long-term evolution of the quasi-steady thermal
luminosities for different neutron star masses (coded with colors and
labeled near the curves) and different average mass
accretion rates (shown with different line styles, according
to the legend).
}
\label{fig:nsgtm}
\end{figure}

Figure \ref{fig:nsgtm} illustrates the influence of neutron star mass
$M$ and average accretion rate $\langle\dot{M}\rangle$ on the long-term
evolution of the thermal luminosity in quiescence. It is computed for
the basic model with the accreted FZCPHG crust, which gradually replaces the
ground-state BSk24 crust. The direct Urca processes are forbidden for
the two lower masses shown in the figure and open for the two higher
masses. Accordingly, these massive stars cool down quickly via neutrino
emission and have smaller thermal photon luminosities.

We see that, under the assumption of constant average accretion rate,
the long-term evolution of the quiescent thermal luminosity is
non-monotonous. After initial cooling, it has a minimum and then
increases due to continued accumulation of the accreted matter and
activation of deeper reaction shells in the crust. In reality, the
accretion rate can vary. For instance, accretion can start after a long
period of pure cooling. In this case, the minimum of the
luminosity can be much lower than shown in our figures. 

\begin{figure}
\centering
\includegraphics[width=\columnwidth]{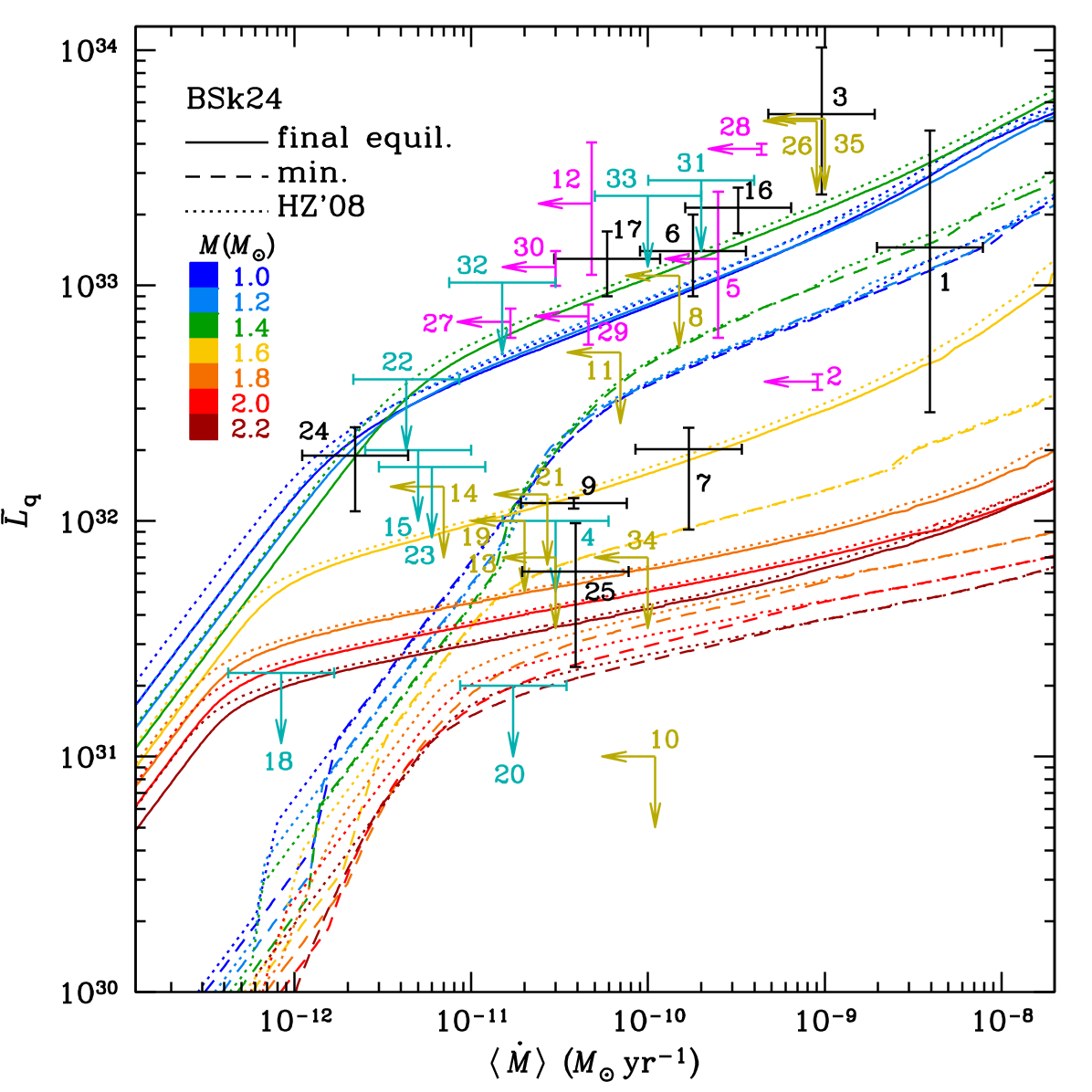}
\caption{Quasi-equilibrium redshifted luminosities as functions of the
average mass accretion rate $\langle\dot{M}\rangle$ for different
neutron star masses (coded with color) in the basic neutron-star model.
Solid lines show the maximum luminosity $\tilde{L}_\mathrm{q}$
for the FZCPHG model of deep
crustal heating, dashed lines show the minimum $\tilde{L}_\mathrm{q}$
for the same model, and dotted lines display the results for the HZ'08
accreted crust heating model (both maximum and minimum).
}
\label{fig:LvsMdot}
\end{figure}

\begin{figure}
\centering
\includegraphics[width=\columnwidth]{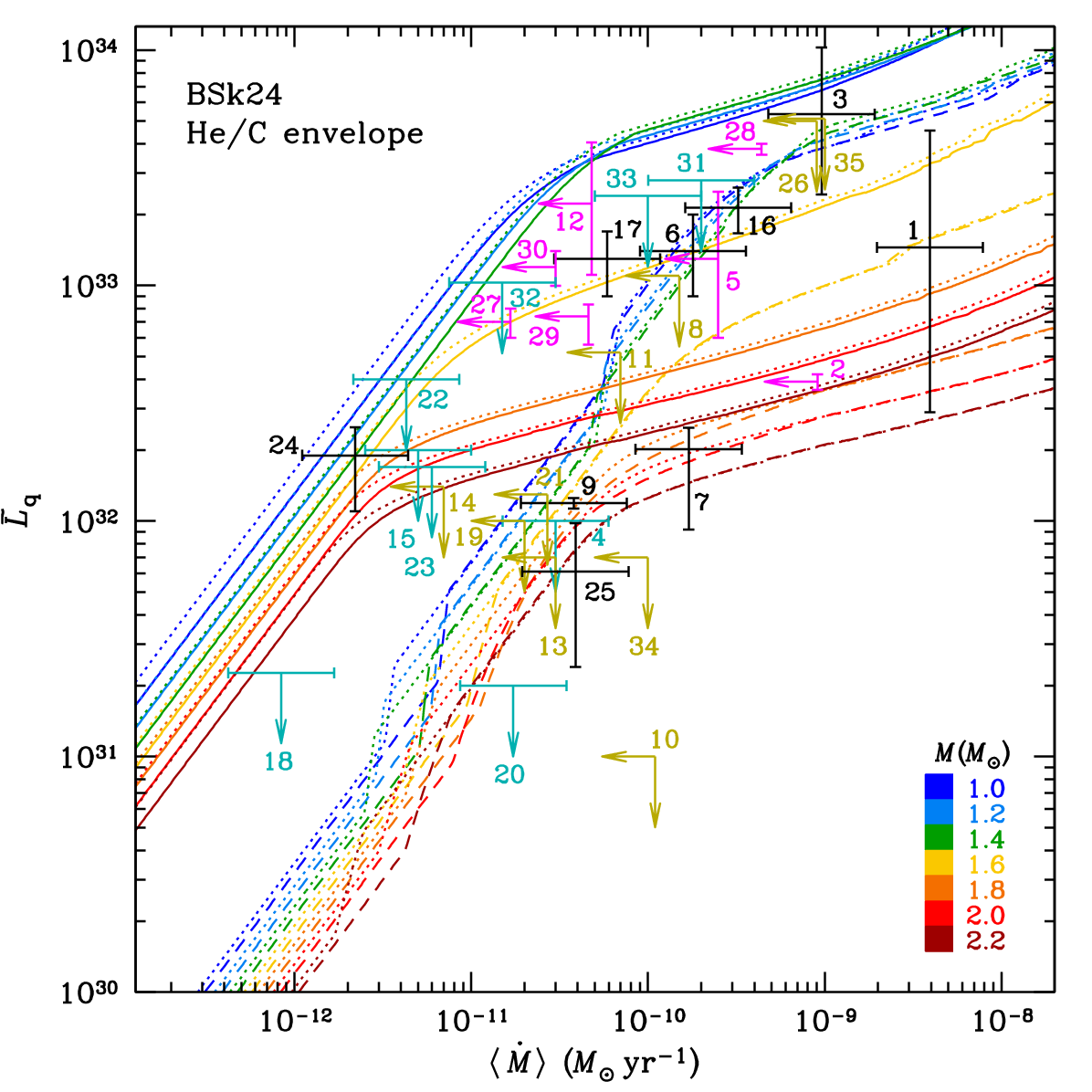}
\caption{The same as in Fig.~\ref{fig:LvsMdot} but for a light-element
composition of the outer envelope from the surface to $\rho=10^9$ \gcc.
}
\label{fig:LvsMdotHeC}
\end{figure}

\begin{figure}
\centering
\includegraphics[width=\columnwidth]{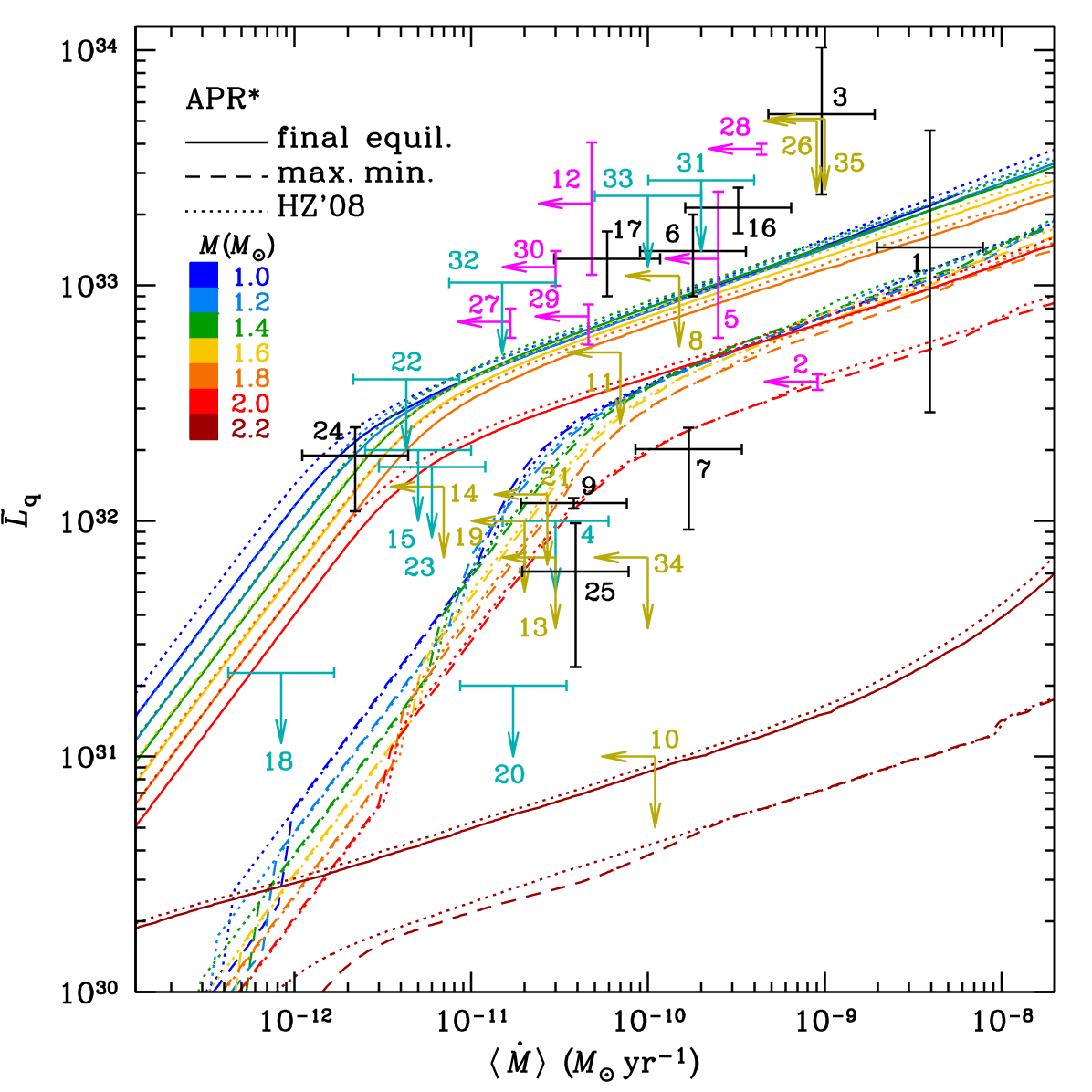}
\caption{The same as in Fig.~\ref{fig:LvsMdot} but for the APR$^*$ EoS,
composition, and effective masses.
}
\label{fig:LvsMdotAPR}
\end{figure}

\begin{figure}
\centering
\includegraphics[width=\columnwidth]{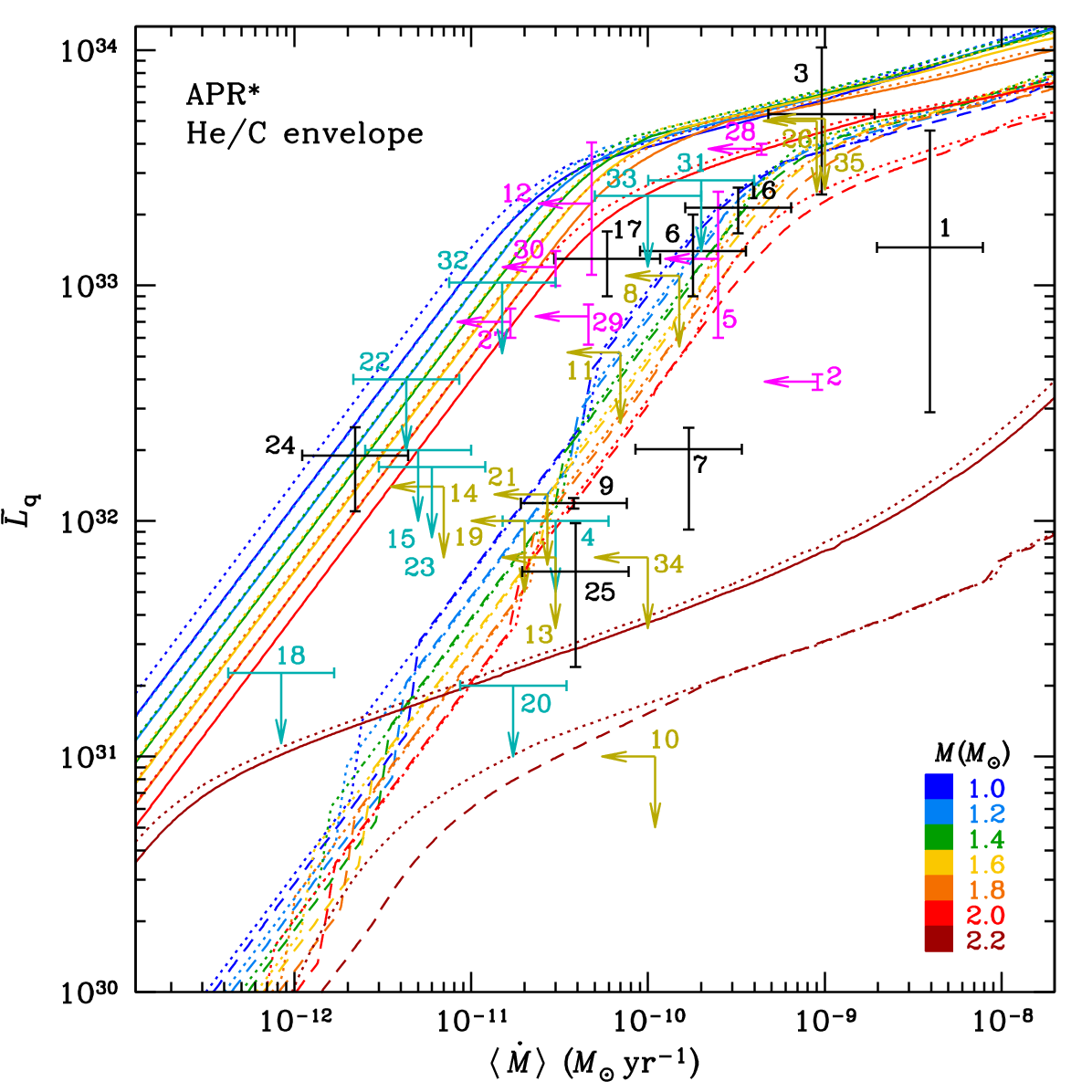}
\caption{The same as in Fig.~\ref{fig:LvsMdotAPR} but with an accreted
envelope as in Fig.~\ref{fig:LvsMdotHeC}.
}
\label{fig:LvsMdotAPRHeC}
\end{figure}

In any case, the quiescent luminosity for a given SXT can have any value
in a ``window'' between the minimum and the final quasi-steady state at
any given average accretion rate. Coming back to the simplest assumption
of  constant $\langle\dot{M}\rangle$, we can calculate this window,
which is sliding as function of $\langle\dot{M}\rangle$. The results of
such calculations are shown in
Figs.~\ref{fig:LvsMdot}\,--\,\ref{fig:LvsMdotAPRHeC} for the BSk24 and
APR$^*$ models of the interior with the traditional iron heat blanketing
envelope and with heat-transparent He/C envelope. We see that the basic
model  (Fig.~\ref{fig:LvsMdot}) is unable to explain the whole range of the estimated values of
$\tilde{L}_\mathrm{q}$ and $\langle\dot{M}\rangle$ simultaneously. At a
given $\langle\dot{M}\rangle$, the hottest objects (numbers 3, 6, 12,
28) are brought to
agreement with the theory by the assumption that their heat-blanketing
envelopes are composed of light elements (Figs.~\ref{fig:LvsMdotHeC},
\ref{fig:LvsMdotAPRHeC}). In addition, the masses of these hot objects
should not exceed $M_\mathrm{DU}$. On the contrary, several
coldest objects (numbers 4, 7, 20, 25) are better explained without the
accreted envelopes and with $M>M_\mathrm{DU}$. 

Object 20 (SAX J1808.4$-$3658) appears to be incompatible with BSk24
model if its crust is entirely replaced by the accreted material.  An
upper bound on the quiescent luminosity of this SXT, evaluated from
observations, restricts $\tilde{L}_\mathrm{q}$ to values that are very
low for its estimated accretion rate. In the APR$^*$ model, it can be
explained (Figs.~\ref{fig:LvsMdotAPR}, \ref{fig:LvsMdotAPRHeC}), but
only if its mass appreciably exceeds $2\,M_\odot$.  However, an analysis
based on evolutionary scenarios \citep{Tailo_ea18} favors $M\sim
1.6\,M_\odot$, which agrees with results of \citet{MorsinkLeahy11}. If
the crust is only partially replaced, so that the luminosity is near the
minimum, then this SXT is compatible with the BSk24 model for any mass,
but only marginally (Figs.~\ref{fig:LvsMdot}, \ref{fig:LvsMdotHeC}).
However, the mass of the donor star, estimated from observations, is
very low, $M_\mathrm{d}\sim0.04$\,--\,0.07,
\citep{Wang_ea13,Sanna_ea16}, which implies a large accreted mass
$\Delta M_\mathrm{acc}\sim0.2\,M_\odot$ \citep{Tailo_ea18}. The short
spin period of this pulsar (2.5 ms) corroborates a large accreted mass
and hence the fully replaced crust. In the next section we will see that
this difficulty is resolved, when one takes recent advances
in the theory of baryon superfluidity into account.

\section{The effect of triplet baryon pairing suppression}
\label{sect:nt_suppr}

The superfluidity is known to affect neutron star thermal evolution (see
Sect.~\ref{sect:SF}). In particular, the powerful direct Urca processes,
being open at $\bar{n}>\bar{n}_\mathrm{DU}$, can be still strongly
suppressed by baryon superfluidity \citep{YKGH}. It is likely that in
the core of a neutron star the proton singlet superfluidity is the
strongest one (has the highest critical temperature), but only up to a
density of $\rho\sim(3-5)\times10^{14}$ \gcc{}
(Fig.~\ref{fig:meff_Tc_Y}). At higher densities, the neutron triplet
superfluidity comes into play. These higher densities are most important
for the neutrino emission by the direct Urca process. As a rule, the
neutron triplet superfluidity has a lower critical temperature than the
proton singlet one. Note that even if the maximum triplet pairing gap is
similar to that of the singlet type, $T_\mathrm{crit}$ still is lower by
a factor of $\approx1/5$ due to the anisotropy of the triplet gap (e.g.,
\citealt{AmundsenOstgaard85,Baldo_ea92}; cf.~\citealt{Ho_ea15}; this
factor of difference is sometimes overlooked in the literature on
neutron star cooling). However, according to several recent studies (see
\citealt{SedrakianClark} for review and references), many-particle
correlations in the baryon matter lead to strong suppression of the
triplet type of superfluidity. In particular, the results of
\citet{Ding_ea16} suggest that the pairing gap may be reduced by an
order of magnitude or even stronger at high densities,  as illustrated
in the middle panel of Fig.~\ref{fig:meff_Tc_Y}. To test the
effect of this suppression, we compare the neutron star cooling
and heating  for the nt-type superfluidity models ``Av18 SRC+P'' and
``N3LO SRC+P'' by \citet{Ding_ea16} with the BEEHS model by
\citet{Baldo_ea98}.

\begin{figure}
\centering
\includegraphics[width=\columnwidth]{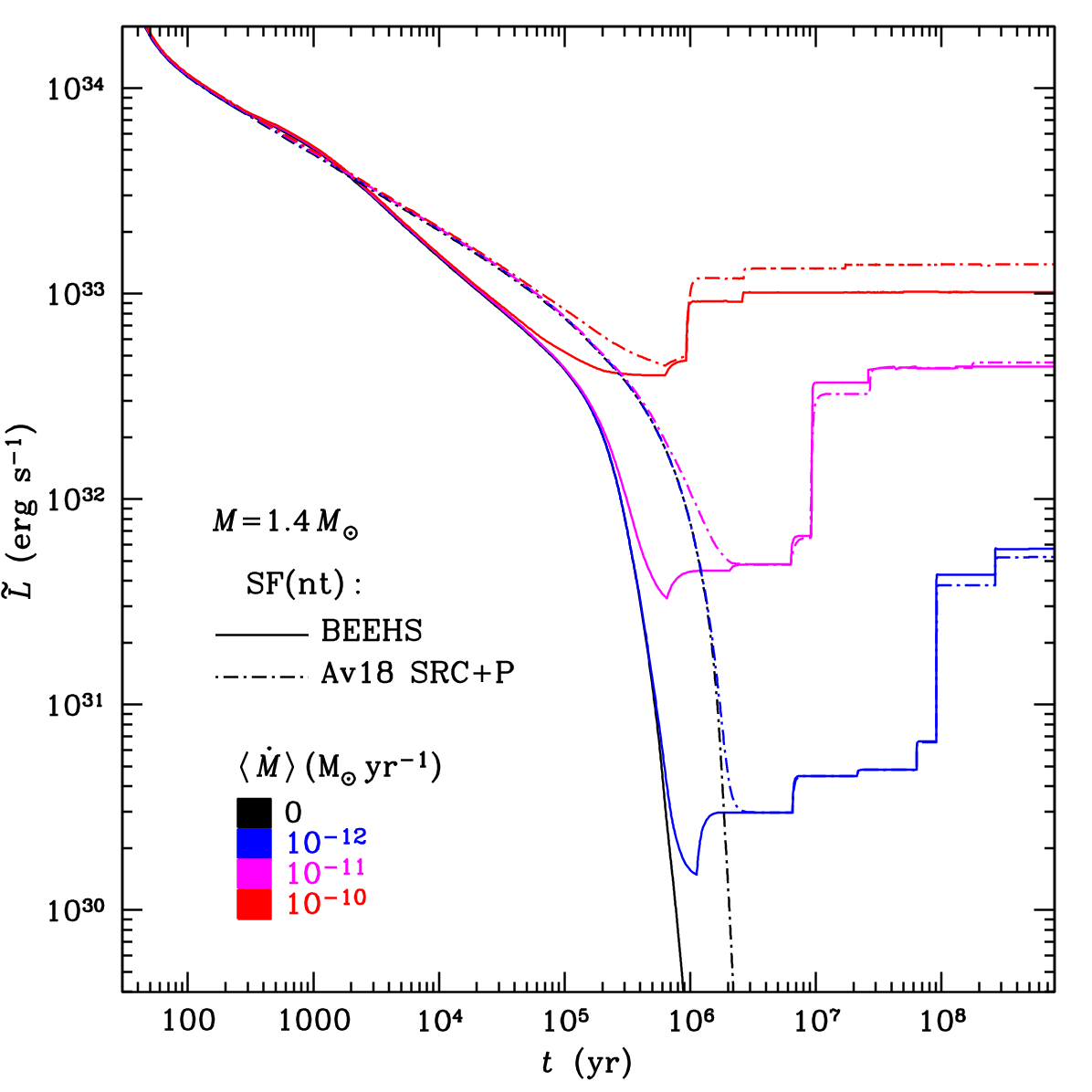}
\caption{Long-term evolution of the quasi-steady thermal
luminosities for the basic model of a neutron star ($M=1.4\,M_\odot$,
standard superfluidity) without accretion and
with long-term steady accretion at different
rates $\langle\dot{M}\rangle$ (solid lines), 
compared with evolution of the same
star but with suppressed neutron triplet superfluidity in the core
(dot-dashed curves) according to the ``Av18 SRC+P'' pairing gap model
of \citet{Ding_ea16}.
}
\label{fig:nsgtsfd}
\end{figure}

The results of such simulations for a neutron star with $M=1.4\,M_\odot$
are shown in Fig.~\ref{fig:nsgtsfd}. We see that the suppression of the
nt-superfluidity delays cooling at the late time of evolution and
increases  thermal luminosity at sufficiently high average accretion
rates.  The PBF neutrino emission is most powerful at $T\sim
T_\mathrm{crit}$ and is effectively quenched by the decrease of
$T_\mathrm{crit}$.

\begin{figure}
\centering
\includegraphics[width=\columnwidth]{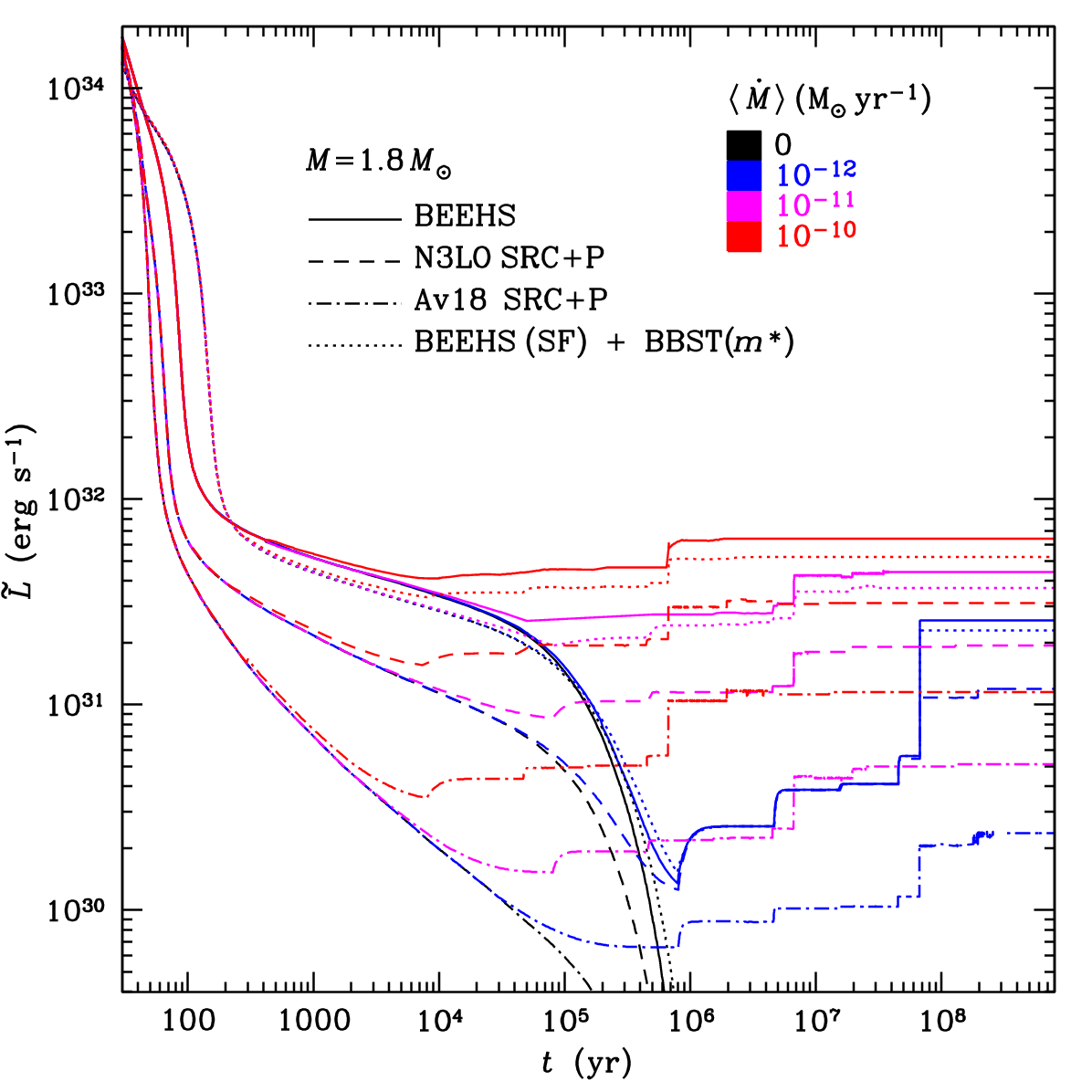}
\caption{Long-term evolution of the quasi-steady thermal luminosities
for a neutron star with $M=1.8\,M_\odot$ without accretion and with
long-term steady accretion at different rates $\langle\dot{M}\rangle$
for different physics inputs in the core. The model with the standard
nucleon superfluidity and  effective masses consistent with the BSk24
EoS (solid lines) is compared with the results of using suppressed
neutron triplet superfluidity  in the core according to models of
\citet{Ding_ea16} ``N3LO SRC+P'' (dashed lines) and ``Av18 SRC+P''
(dot-dashed lines), or with the alternative model (BBST instead of
BSk24) for the nucleon effective masses (dotted curves).
}
\label{fig:nsgtM1.8}
\end{figure}

Analogous comparison for a more massive neutron star is presented in
Fig.~\ref{fig:nsgtM1.8}. The principal difference of this case from the
previous one is that the direct Urca process operates in such a star. We
see that the suppression of the nt-superfluidity for the massive star
has an opposite effect compared to the previous figure: the cooling is
accelerated, and the heating phase shows lower luminosities. The reason
is that the direct Urca emission fades away when baryon superfluidity
develops. When superfluidity is partially suppressed, the direct Urca
process partially regains its power. For comparison, in
Fig.~\ref{fig:nsgtM1.8} we also show the thermal evolution computed with
an alternative model of nucleon effective masses (BBST instead of BSk24,
see Fig.~\ref{fig:meff_Tc_Y}). Since the effective masses affect the
neutrino reaction rates, we see some differences at the stages of
thermal evolution where the neutron star interior is sufficiently hot,
so that the neutrino energy losses dominate over conductive losses.
However, the dependence on  the effective masses is seen to have a much
smaller effect than the dependence on the superfluidity. 

\begin{figure}
\centering
\includegraphics[width=\columnwidth]{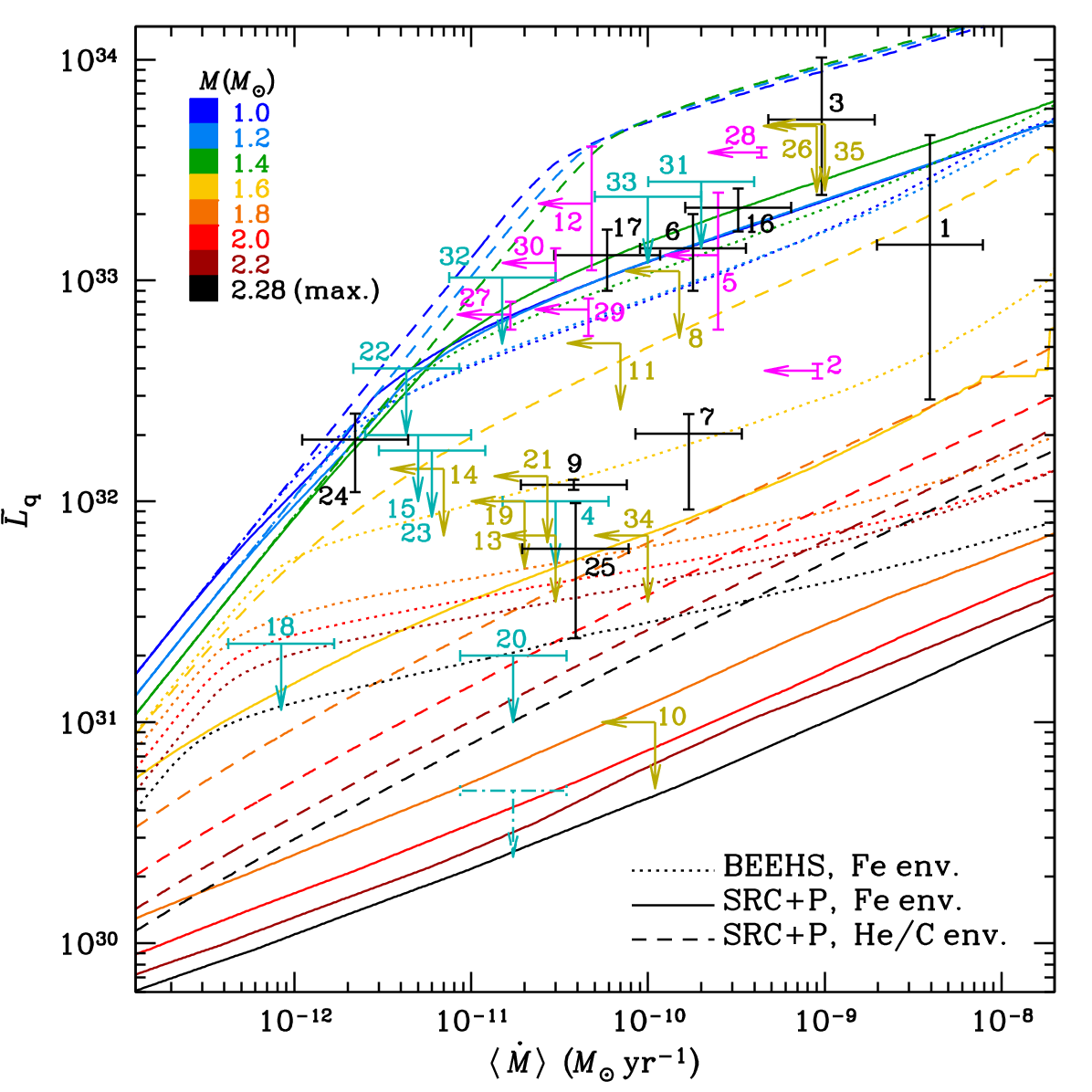}
\caption{Quasi-equilibrium redshifted luminosities as functions of the
average observed mass accretion rate
$\langle\dot{M}_\mathrm{obs}\rangle$, computed including the effect of
suppression of the neutron triplet superfluidity. Solid lines are
obtained for the iron heat-blanketing envelope and dashed lines for the
accreted (He/C) envelope. For comparison, the dotted lines show the
luminosities with non-suppressed triplet superfluidity, as  in
Fig.~\ref{fig:LvsMdot}. The solid errorbars and arrows show the
estimated values and upper limits listed in Table~\ref{tab:obj},  as in
the previous figures, and the additional dot-dashed ones show the
traditional tightest estimate of an upper bound on the thermal
luminosity of SAX J1808.4$-$3658 in quiescence, but with the updated
accretion rate.
}
\label{fig:LvsMdotSFD}
\end{figure}

The heating curves for neutron stars of different masses with  the
partly suppressed nt-superfluidity are shown in
Fig.~\ref{fig:LvsMdotSFD} for the cases with iron and light-element heat
blanketing envelopes. For comparison we also show the heating curves
obtained with the non-suppressed nt-superfluidity. We see that the
suppression of the nt-superfluidity decreases the smaller thermal
luminosities, appropriate to most massive neutron stars, and increases
the higher luminosities, appropriate to neutron stars with lower masses.
The first effect is explained by the fact that the direct Urca processes
are strongly suppressed at $T\ll T_\mathrm{crit}$. When $T_\mathrm{crit}$
decreases, these processes take away more energy from the core and thus
cool down the neutron star more efficiently. The second effect is due to
the PBF mechanism. This mechanism of neutrino emission is entirely due
to superfluidity, so the partial suppression of superfluidity delays the
PBF processes and thus preserves more heat inside the star. Thus the
lower heating curves are pushed downward and the higher upward, which
facilitates theoretical interpretation of the low and high values of
$\tilde{L}_\mathrm{q}$.  In particular, in this way we can explain not
only the conservative upper limit on the quiescent thermal luminosity of
one of the coldest transiently accreting neutron stars in SAX
J1808.4$-$3658,  but also the tightest, non-conservative limit,
traditionally used in the literature
\citep[e.g.,][]{BeznogovYakovlev15a}, displayed against the updated
estimate of the average accretion rate \citep{CoriatFD12,VanIH19}.

The hottest neutron stars in the SXTs still
remain only marginally compatible with theoretical heating curves,
computed in the models with iron heat blanketing envelopes. But an
inclusion of an accreted outer envelope into the model increases the
observed luminosities and thereby provides an easy explanation of all
simultaneous estimates of $\tilde{L}_\mathrm{q}$ and
$\langle\dot{M}_\mathrm{obs}\rangle$ in Table~\ref{tab:obj}.

\section{Conclusions}

We have revisited the evaluation of quasi-equilibrium thermal
luminosities of neutron stars in SXTs in quiescence, taking
the recent progress in observations of the SXTs and in the theory of
neutron stars into account. We have composed an updated collection of the key
properties of SXTs with estimated average mass accretion rates and
neutron-star luminosities in quiescence. We have simulated long-term
thermal evolution and computed thermal states of the SXTs with different
mass accretion rates for different modern theoretical neutron-star
models with the nucleon-lepton ($npe\mu$) composition of the core. We
explored the possibility that the neutron-star crust is not
completely replaced by the reprocessed accreted matter. In particular,
we have computed the minimal quiescent thermal luminosities in the
simplest model of an accretion at constant average rate. In this
model, the minimal theoretical luminosity of the transiently accreting
neutron stars that are relatively cool for their estimated average
accretion rates becomes compatible with theory (albeit marginally) even
without invoking the enhanced (direct Urca) cooling. However,  their
short spin periods ($1.8$\,--\,2.7~ms) suggest a large accreted mass and
therefore disfavor such a scenario. Indeed, the  evolutionary
life-time of a LMXB (gigayears) is longer than the time needed to
accumulate the accreted mass $\Delta
M_\mathrm{acc}\gtrsim0.002\,M_\odot$ that is sufficient to
reach the steady quiescent equilibrium, and orders of magnitude longer
than a megayear when the quiescent luminosity is at
minimum. Therefore, the number of LMXB systems with neutron stars in this
minimum state should be relatively small.

On the other hand, there are several transiently accreting neutron stars
that are relatively hot for their estimated accretion rates, which can
be explained by the presence of a relatively heat-transparent accreted
outer envelope. Thus the updated observational data and updated
theoretical physics input leave unchanged the basic conclusions of
\citet{YakovlevLH03,Yakovlev_ea04} on the possible explanations of the
hottest and coldest neutron stars in SXTs. The replacement of the
accreted crust model HZ'08 \citep{HZ08} by the new model FZCPHG
\citep{Fantina_ea18} does not change any qualitative conclusions.

One of the coldest neutron stars in SXTs, SAX J1808.4$-$3658, still present
a difficulty for theoretical interpretation. We have found that the
difficulty is related to the suppression of the direct Urca process by
neutron triplet superfluidity in the core.  Allowance for quenching of
this type of superfluidity according to the results of \citet{Ding_ea16}
makes the theoretical heating curves of this and other relatively cold
transiently accreting neutron stars fully compatible with observations.
Moreover, the same quenching brings the theory to better agreement with
observed thermal luminosities of relatively hot transiently accreting
neutron stars in quiescence, although the presence of an accreted
envelope is still needed for such agreement. Thus the observational data
on neutron-star heating in SXTs favor the suppression of the neutron
triplet type of superfluidity, which is in line with the analogous
conclusions made recently for the isolated cooling neutron stars
\citep{Beznogov_ea18, WeiBS19}. 

\begin{acknowledgements}

We are grateful to A.~D.~Kaminker and P.~S.~Shternin for fruitful
discussions and to P.~Haensel and J.~L.~Zdunik for useful communications
and for providing some details of their models of the accreted
neutron-star crusts. The work of A.C.{} and A.P.{} was supported  by the
Russian Science Foundation (grant 19-12-00133).

\end{acknowledgements}

\newcommand{\artref}[4]{{#4}, {#1}, {#2}, #3}
\newcommand{\AandA}[3]{\artref{A\&A}{#1}{#2}{#3}}
\newcommand{\AJ}[3]{\artref{\aj}{#1}{#2}{#3}}
\newcommand{\ApJ}[3]{\artref{\apj}{#1}{#2}{#3}}
\newcommand{\ApJS}[3]{\artref{\apjs}{#1}{#2}{#3}}
\newcommand{\ApSS}[3]{\artref{\apss}{#1}{#2}{#3}}
\newcommand{\ARAA}[3]{\artref{\araa}{#1}{#2}{#3}}
\newcommand{\MNRAS}[3]{\artref{\mnras}{#1}{#2}{#3}}
\newcommand{\NAR}[3]{\artref{\nar}{#1}{#2}{#3}}
\newcommand{\NP}[4]{\artref{Nucl.\ Phys. #1}{#2}{#3}{#4}}
\newcommand{\PL}[4]{\artref{Phys.\ Lett. #1}{#2}{#3}{#4}}
\newcommand{\PR}[4]{\artref{Phys.\ Rev. #1}{#2}{#3}{#4}}
\newcommand{\PRL}[3]{\artref{\prl}{#1}{#2}{#3}}
\newcommand{\RMP}[3]{\artref{Rev.\ Mod.\ Phys.}{#1}{#2}{#3}}
\newcommand{\SSRv}[3]{\artref{\ssr}{#1}{#2}{#3}}

\end{document}